\documentclass[journal,twocolumn]{IEEEtran}
\usepackage{graphicx,times,amsmath,amssymb,cite,subfigure,stfloats,booktabs, url,multirow,afterpage}
\usepackage[lined,ruled]{algorithm2e}
\allowdisplaybreaks

\usepackage{bbding}
\usepackage[normalem]{ulem}
\useunder{\uline}{\ul}{}

\makeatletter

\newcommand{\Rmnum}[1]{\expandafter\@slowromancap\romannumeral #1@}
\makeatother

\begin{document}
\title{SAFE: Secure and Accurate Federated Learning for Privacy-Preserving Brain-Computer Interfaces}

\author{Tianwang~Jia, Xiaoqing~Chen and Dongrui~Wu
\thanks{T. Jia, X. Chen and D. Wu are with the Key Laboratory of the Ministry of Education for Image Processing and Intelligent Control, School of Artificial Intelligence and Automation, Huazhong University of Science and Technology, Wuhan 430074, China. X. Chen and D. Wu are also with Zhongguancun Academy, Beijing, 100084 China. Email: \{twjia, xqchen914, drwu\}@hust.edu.cn.}
\thanks{This research was supported by Open Research Fund of the State Key Laboratory of Brain-Machine Intelligence, Zhejiang University (Grant No. BMI2400015).}
\thanks{Dongrui~Wu is the corresponding author.  Email: drwu09@gmail.com.}}

\maketitle

\begin{abstract}
Electroencephalogram (EEG)-based brain-computer interfaces (BCIs) are widely adopted due to their efficiency and portability; however, their decoding algorithms still face multiple challenges, including inadequate generalization, adversarial vulnerability, and privacy leakage. This paper proposes Secure and Accurate FEderated learning (SAFE), a federated learning-based approach that protects user privacy by keeping data local during model training. SAFE employs local batch-specific normalization to mitigate cross-subject feature distribution shifts and hence improves model generalization. It further enhances adversarial robustness by introducing perturbations in both the input space and the parameter space through federated adversarial training and adversarial weight perturbation. Experiments on five EEG datasets from motor imagery (MI) and event-related potential (ERP) BCI paradigms demonstrated that SAFE consistently outperformed 14 state-of-the-art approaches in both decoding accuracy and adversarial robustness, while ensuring privacy protection. Notably, it even outperformed centralized training approaches that do not consider privacy protection at all. To our knowledge, SAFE is the first algorithm to simultaneously achieve high decoding accuracy, strong adversarial robustness, and reliable privacy protection without using any calibration data from the target subject, making it highly desirable for real-world BCIs.
\end{abstract}


\IEEEpeerreviewmaketitle

\section{Introduction}

A brain-computer interface (BCI)\cite{Wolpaw2002Brain} enables direct communication between the brain and external devices, offering applications in neurological rehabilitation\cite{Lorach2023Walking}, robot control~\cite{Hochberg2012Reach}, text input \cite{Chen2015High,Willett2021High}, speech synthesis~\cite{Metzger2023high}, affect recognition and regularization \cite{Wu2023Affective}, and so on. Electroencephalography (EEG), due to its efficiency and portability, is one of the most commonly used non-invasive BCI inputs. Among various EEG-based BCI paradigms, motor imagery (MI)~\cite{Pfurtscheller2001Motor}, event-related potentials (ERPs)~\cite{Hoffmann2008efficient}, and steady-state visual evoked potentials (SSVEPs)~\cite{Friman2007multiple}, are widely adopted.

Despite the advantages of EEG-based BCIs, they also face multiple challenges in real-world applications, including:
\begin{enumerate}
\item \emph{Inadequate generalization.} A well-generalizable EEG classifier is critical for real-world BCIs, especially for new users with little or no calibration data. However, conventional EEG classifiers trained for individual subjects or on limited datasets often generalize poorly across users, due to the inherently low signal-to-noise ratio of EEG signals and substantial inter-subject variations~\cite{Wu2022Transfer}.

\item \emph{Adversarial vulnerability.} BCI systems exhibit serious vulnerabilities to adversarial attacks, where malicious small perturbations are injected into EEG data to induce erroneous classification outputs, as illustrated in Fig.~\ref{fig:attack_bci}. Recent studies have demonstrated that even tiny adversarial perturbations can significantly compromise the decoding performance~\cite{Zhang2019vulnerability,Zhang2021tiny,Jung2023generative,Meng2024adversarial}. Adversarial attacks pose significant risks in critical BCI applications, as they may result in incorrect clinical decisions in healthcare, and dangerous operations in military operations \cite{Binnendijk2020Brain}.

\begin{figure}[htbp]     \centering
	\includegraphics[width=.9\linewidth]{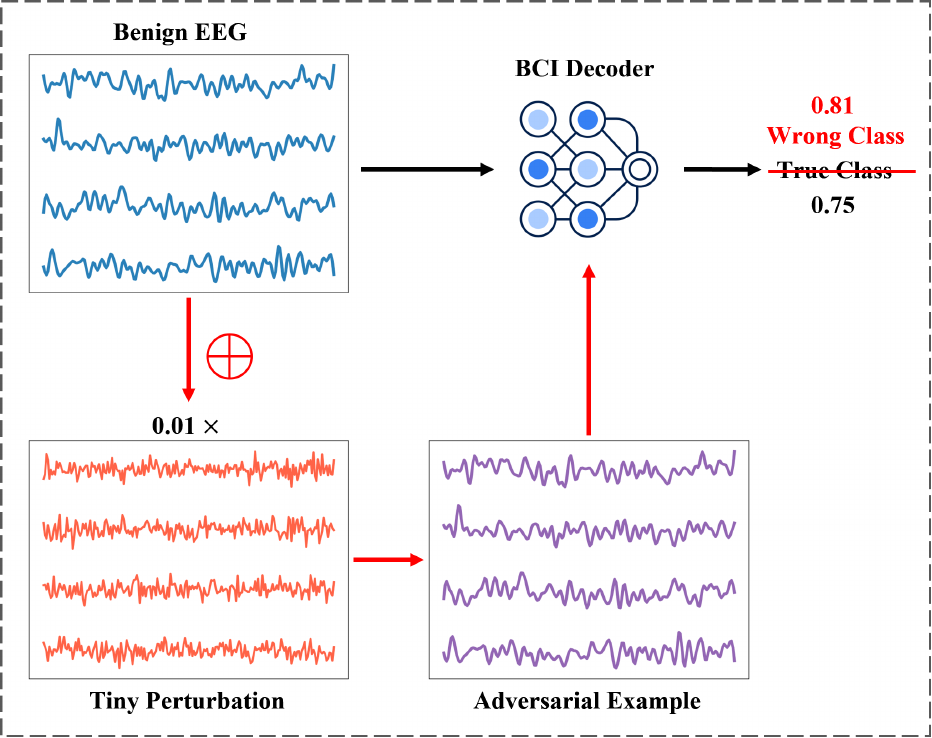}
	\caption{Adversarial attacks on BCIs.}     \label{fig:attack_bci}
\end{figure}

\item \emph{Privacy leakage.} Regulations such as European Union's General Data Protection Regulation and China's Personal Information Protection Law enforce strict protocols for collecting, storing and utilizing personal sensitive data. Previous studies have shown that EEGs can expose user privacies like mental diseases and cognitive states~\cite{Xia2023Privacy}. Thus, privacy-preserving BCIs become a necessity.
\end{enumerate}

Substantial progresses have been made to address each individual challenge of the above, including decoding accuracy enhancement~\cite{He2020Transfer,Wu2022Transfer,Wu2025Revisiting}, adversarial robustness~\cite{Meng2023Adversarial, Chen2024Adversarial,Chen2025Data}, and privacy protection~\cite{Meng2023User,Shao2024Machine,Chen2025User}. Several studies have attempted to tackle two of the three challenges simultaneously. For instance, Zhang \emph{et al.}~\cite{Zhang2022Lightweight} considered simultaneously the decoding accuracy and privacy protection, and Chen \emph{et al.}~\cite{Chen2024Alignment} enhanced both the decoding accuracy and adversarial robustness.

\begin{figure}[htbp]     \centering
	\includegraphics[width=0.8\linewidth]{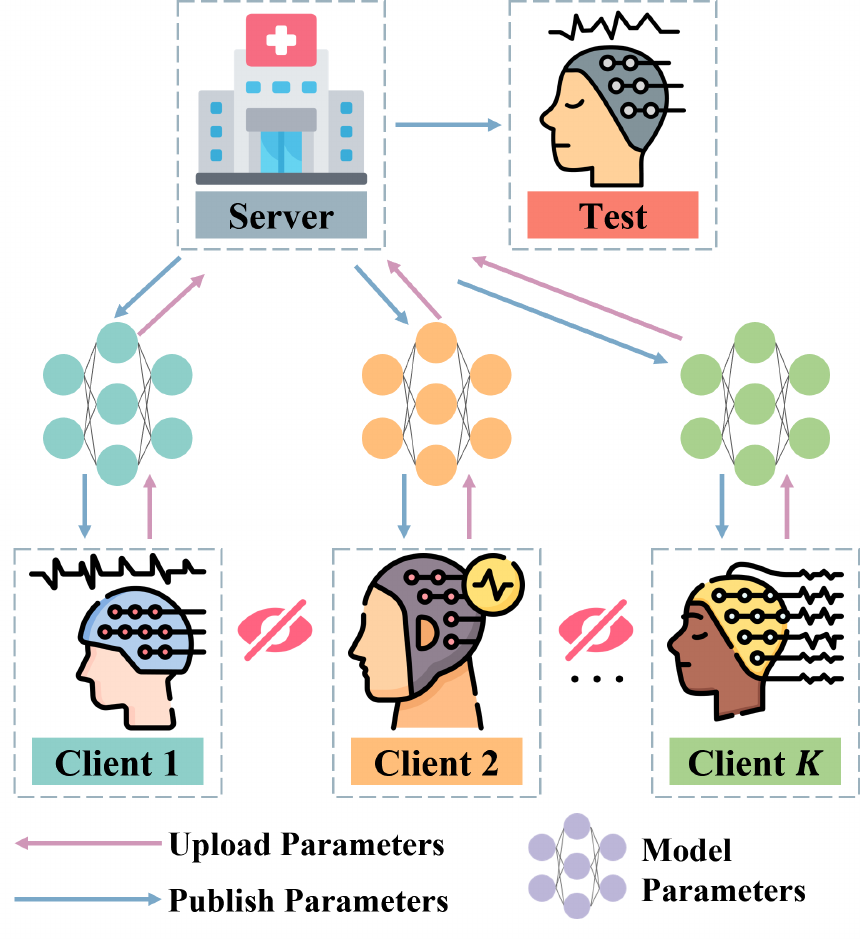}
	\caption{SAFE for BCIs. Training users are clients, and a trusted third-party institution (e.g., a hospital) is the server; only model parameters, instead of EEG data, are exchanged, protecting client user privacy.}     \label{fig:safe_for_bci}
\end{figure}

To our knowledge, Chen \emph{et al.}~\cite{Chen2024Accurate} were the first to demonstrate the feasibility of addressing all three challenges simultaneously. However, their approach needs to make use of some target domain calibration data. This paper proposes Secure and Accurate FEderated learning (SAFE), which is completely calibration-free for the new subject. It builds upon a federated learning (FL)~\cite{Yang2019Federated} architecture, as shown in Fig.~\ref{fig:safe_for_bci}. Each client retains its own EEG data, trains a local copy of the classifier, and uploads the model parameters to the central server. The central server then aggregates the model parameters from all clients and publishes them to the clients again for updates. In this way, a global EEG classifier can be trained without exposing any client's EEG data to any other clients or the test user, ensuring strong privacy protection.

We conducted experiments on five EEG datasets from MI and ERP paradigms to compare SAFE with 14 state-of-the-art approaches, including seven FL methods with privacy protection and seven centralized training (CT) methods without any privacy protection. We evaluated SAFE on benign data and adversarial examples generated by five representative black-box/white-box adversarial attack approaches. The results demonstrated that SAFE significantly improved the decoding accuracy and adversarial robustness while ensuring privacy protection.

The remainder of this paper is organized as follows. Section~\ref{sec:related} reviews related work in domain generalization, adversarial defense, and privacy-preserving machine learning. Section~\ref{sec:safe} proposes SAFE. Section~\ref{sec:settings} describes the experiment setup. Section~\ref{sec:results} presents the experimental results on five datasets. Finally, Section~\ref{sec:conclusions} draws conclusions.

\section{Related Work} \label{sec:related}

This section reviews related work addressing three key challenges in machine learning: inadequate generalization, adversarial vulnerability, and privacy leakage. Specifically, we focus on studies in domain generalization, adversarial defense, and privacy-preserving machine learning.

\subsection{Domain Generalization}

Domain generalization, also known as out-of-distribution generalization, addresses the challenge of learning a model from one or multiple related source domains that can generalize effectively to unseen target domains~\cite{Zhou2022Domain}. Over the years, significant progress has been made in domain generalization research. Existing approaches can be broadly categorized into three groups: data manipulation, representation learning, and learning strategy~\cite{Wang2022Generalizing}.

Data manipulation aims to improve generalization by transforming the model inputs, primarily through data augmentation~\cite{Shorten2019survey} and data generation~\cite{Zhou2020Learning}.

Representation learning is a commonly used approach in domain generalization, achieved by extracting domain-invariant features across different domains~\cite{Ganin2016Domain,Arjovsky2019Invariant}, or decomposing features into domain-shared and domain-specific components to enhance generalization~\cite{Ding2017Deep}.

Learning strategy refers to the utilization of general learning strategies, such as meta-learning~\cite{Li2018Learning}, gradient manipulation~\cite{Huang2020Self}, distributionally robust optimization~\cite{Sagawa2020Distributionally}, and self-supervised learning~\cite{Carlucci2019Domain}, to improve generalization.

\subsection{Adversarial Defense}

Adversarial defenses have been extensively explored in both image and natural language processing domains to enhance model robustness against adversarial attacks. One of the most prevalent approaches is adversarial training (AT)~\cite{Madry2018Towards}, which augments the benign training data with adversarial samples. Other commonly adopted methods include model-level~\cite{Addepalli2020Towards,Folz2020Adversarial} and data-level~\cite{Guo2018Countering,Jia2019Comdefend} defenses.

Zhang \emph{et al.}~\cite{Zhang2019vulnerability} first discovered that adversarial samples significantly reduce the decoding accuracy of EEG-based BCIs. Subsequently, multiple adversarial defense approaches for BCIs have been proposed. Meng \emph{et al.}~\cite{Meng2023Adversarial} systematically evaluated the advantages and limitations of various defense strategies in BCI applications. Gunawardena \emph{et al.}~\cite{Gunawardena2024Single} combined AT with adversarial detection, effectively enhancing the robustness of BCIs against black-box attacks. Chen \emph{et al.}~\cite{Chen2024Alignment} proposed ABAT, which performs AT on the aligned EEG data to improve the decoding accuracies on both benign and adversarial samples.

\subsection{Privacy-preserving Machine Learning}

Privacy-preserving machine learning avoids using the raw data directly, protecting their privacy. Key methodologies include source-free domain adaptation and FL. 

Source-free domain adaptation protects data privacy by transferring knowledge from the source domain to the target domain without accessing the source domain data. For example, SHOT~\cite{Liang2020Do} utilizes only the source model and unlabeled target data in pseudo-label-based self-supervised adaptation. 

FL establishes a privacy-preserving distributed learning framework, where clients collaboratively train models without sharing raw data. FedAvg~\cite{McMahan2017Communication}, the most prevalent FL approach, reduces communication overhead by conducting multiple local stochastic gradient descent updates per client before server-side model aggregation. Many studies have improved FedAvg to address various challenges, such as enhancing adversarial robustness~\cite{Li2023Federated}, mitigating client data drift~\cite{Li2020Federated,Karimireddy2020Scaffold,Li2021Model}, and improving model generalization~\cite{Zhou2023FedFA,Zhang2023Federated}.

Privacy protection is particularly critical for BCI applications, since commercial BCI systems often require cross-institutional collaboration (e.g., hospitals, universities, and companies), and EEG data inherently contain sensitive private information. Different approaches for privacy-preserving BCIs have been proposed. Xia \emph{et al.}~\cite{Xia2022Privacy} proposed augmentation-based source-free domain adaptation for cross-subject MI classification. Zhang \emph{et al.} proposed lightweight source-free transfer~\cite{Zhang2022Lightweight} and multi-source decentralized transfer~\cite{Zhang2022Multi}. FL-based solutions include Ju \emph{et al.}'s~\cite{Ju2020Federated} federated transfer learning framework for extracting shared EEG features, and Liu \emph{et al.}'s~\cite{Liu2024Aggregating} split-classifier architecture combining global-local knowledge. FedBS~\cite{Jia2024Federated} improves both privacy protection and the decoding accuracy, without considering adversarial robustness.

\section{Secure and Accurate FEderated learning (SAFE)} \label{sec:safe}

This section introduces the details of our proposed SAFE approach.

\subsection{Problem Statement} \label{sec:problem}

Assume there are $K$ subjects $\mathcal{S}=\{\mathcal{S}_k\}_{k=1}^K$, each as a client participating in the training. For the $k$-th client, there are $n_k$ labeled EEG samples $\mathcal{S}_k=\{(\mathbf{X}_{k, i}, y_{k,i})\}^{n_k}_{i=1}$ and a classifier with parameters $\boldsymbol\theta_k^r$, where $\mathbf{X}_{k, i}\in \mathbb{R}^{c\times t}$ is the $i$-th EEG trial, $y_{k, i}\in\{1,\cdots,L\}$ is the corresponding label, and $r\in \{1,\cdots, R\}$ is the index of communication rounds, in which $c$, $t$, $L$ and $R$ represent the number of channels, time domain samples, classes, and maximum communication rounds, respectively. The goal of SAFE is to train a well-generalizable EEG classifier without any calibration data from test subjects.

\subsection{Overview}

\begin{figure}[htbp] 	\centering
	\subfigure[Server-side operations in SAFE.]{\label{fig:safe_server}\includegraphics[width=\linewidth]{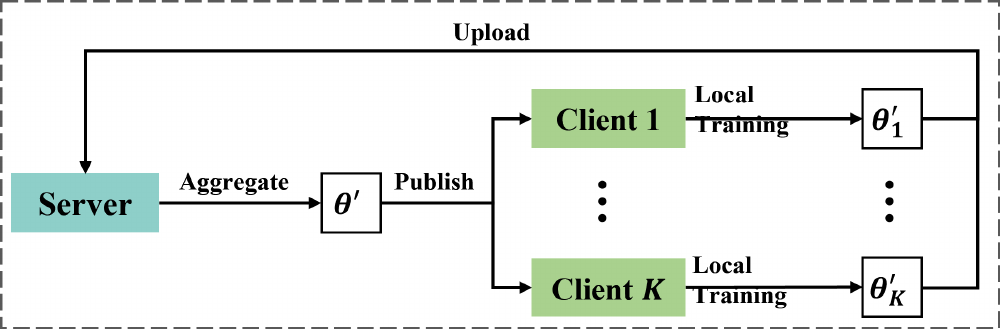}}
	\subfigure[Client-side operations in SAFE.]{\label{fig:safe_client}\includegraphics[width=\linewidth]{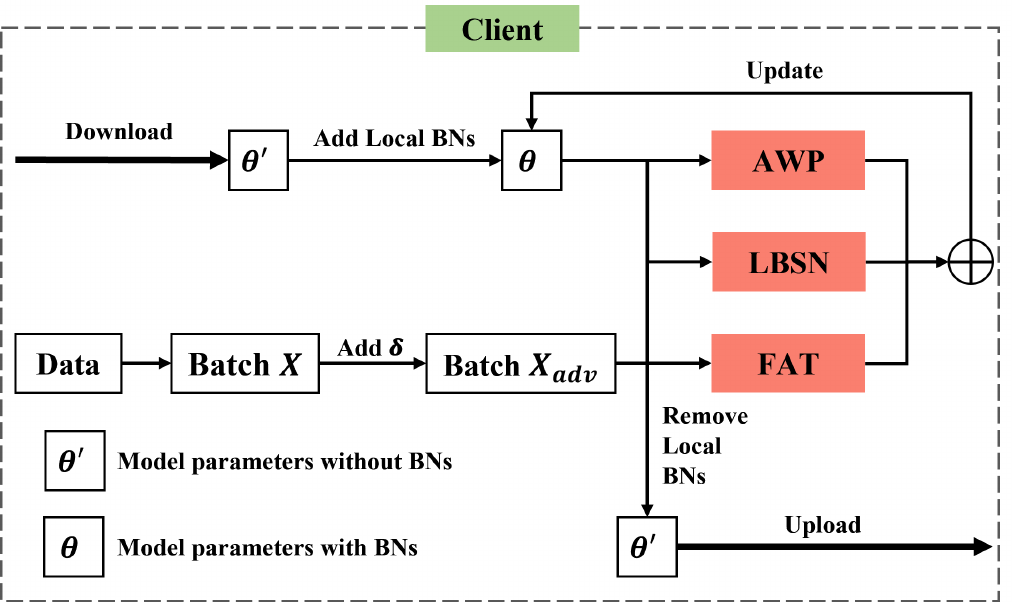}}	
	\caption{Flowchart of the proposed SAFE. (a) Server-side operations; and, (b) client-side operations.} 	\label{fig:safe}
\end{figure}

Fig.~\ref{fig:safe} illustrates the complete workflow of SAFE, which consists of one central server and multiple clients. 

Each client holds EEG data from an individual subject for localized model training, while the central server communicates with the clients and aggregates their uploaded model parameters into a final global model. We utilize local batch-specific normalization (LBSN) to keep BN layers localized to each client without parameter sharing during training, enhancing privacy protection and mitigating inter-subject distribution discrepancies in both training and testing phases. Furthermore, we introduce a dual-defense mechanism into client-side training, i.e., federated adversarial training (FAT) in the input space and adversarial weight perturbation (AWP) in the parameter space, to collaboratively defend against adversarial attacks.

Algorithm~\ref{algo:safe} gives the pseudo-code of SAFE. The source code will be made available after this paper is accepted.

\begin{algorithm}[htbp] \label{algo:safe}
	\KwIn{$K$, number of clients\;
		\hspace*{9mm} $\mathcal{S}_k$, labeled dataset of the $k$-th client\;
		\hspace*{9mm} $R$, maximum communication rounds\;
		\hspace*{9mm} $\eta$, learning rate\;
		\hspace*{9mm} $m$, number of selected clients per round\;
		\hspace*{9mm} $B$, batch size\;
		\hspace*{9mm} $E$, local optimization epochs per client\;
		\hspace*{9mm} $\alpha$, FAT adversarial example magnitude in (\ref{eq:fgsm})\;
		\hspace*{9mm} $\xi$, perturbation scale for AWP in (\ref{eq:nu}).}
	\KwOut{$\boldsymbol\theta$, model parameters with BN layers.}
	~\\
	\textbf{Server executes:}\\
	Initialize $\boldsymbol\theta'_0$, model parameters without BN layers\;
	\For{$r=1,\dots,R$}{
		$C_r\leftarrow \text{Set of $m$ randomly selected clients}$\;
		\For{client $k \in C_r$}{
			$\boldsymbol\theta'^{k}_{r}\leftarrow \text{ClientUpdate}(k,\boldsymbol\theta'_{r-1})$\;
		}
		$N_r\leftarrow \sum_{k\in C_r}n_k$\;
		\vspace{2pt}
		$\boldsymbol\theta'_{r}\leftarrow \sum_{k\in C_r}\frac{n_k}{N_r} \boldsymbol\theta'^k_{r}$\;
	}
	~\\
	\textbf{ClientUpdate$(k,\boldsymbol\theta')$:}\\
	$\boldsymbol\theta \leftarrow \boldsymbol\theta' + \text{local BN parameters}$\;
	$\mathbb{B}\leftarrow \text{Split $\mathcal{S}_k$ into batches of size $B$}$\;
	\For{$\text{epoch}=1,\dots,E$}{
		\For{$\text{batch }\mathcal{B} \in \mathbb{B}$}{
			Calculate $\boldsymbol\delta$ on $\mathcal{B}$ by (\ref{eq:fgsm});\
			
			Generate adversarial examples from $\mathcal{B}$: $\mathcal{B}'=\{ \mathbf{X}_i+\boldsymbol\delta, y_i\}_{i=1}^B$;\
			
			Calculate $\mu_{\mathcal{B}'}$ on $\mathcal{B}'$ by (\ref{eq:bn_mu});\
			
			Calculate $\sigma_{\mathcal{B}'}$ on $\mathcal{B}'$ by (\ref{eq:bn_sigma});\
			
			Update $\boldsymbol\theta$ with $(\mu_{\mathcal{B}'},\sigma_{\mathcal{B}'})$ on $\mathcal{B}'$ by (\ref{eq:nu}) and (\ref{eq:theta});\
		}
	}
	$\boldsymbol\theta' \leftarrow \boldsymbol\theta - \text{local BN parameters}$\;
	\Return $\boldsymbol\theta'$
	\caption{Secure and Accurate FEderated learning (SAFE).} \label{alg:safe}
\end{algorithm}

\subsection{LBSN}

Following FedBS \cite{Jia2024Federated}, LBSN computes batch-specific statistics for the BN layers, instead of conventional global statistics.

The standard BN layer maintains four sets of parameters: statistical parameters $\mu$ (mean) and $\sigma$ (standard deviation) dynamically computed during forward propagation, and learnable parameters $\gamma$ (scaling factor) and $\beta$ (bias) optimized through backpropagation.

Different from traditional approaches that compute and fix statistical parameters across the entire training process, LBSN dynamically computes batch-wise statistical parameters $\mu$ and $\sigma$ to alleviate the non-stationary nature of EEG signals and inter-subject distribution discrepancies. Specifically, given a training/test batch $\mathcal{B} = \{(\mathbf{X}_i, y_i)\}_{i=1}^B$ with batch size $B$, LBSN computes:
\begin{align}
	\mu_{\mathcal{B}}&= \frac{1}{B}\sum_{i=1}^{B}\mathbf{X}_i,\label{eq:bn_mu}\\
	\sigma^2_{\mathcal{B}}&=  \frac{1}{B}\sum_{i=1}^{B}(\mathbf{X}_i-\mu_{\mathcal{B}})^2.\label{eq:bn_sigma}
\end{align}

Furthermore, LBSN localizes the BN learnable parameters $\gamma$ and $\beta$ to each client during federated training, and aggregates them at the server to form a complete model for testing.

LBSN offers three advantages:
\begin{enumerate}
	\item Isolation of BN parameter interference across clients during training, enabling better feature alignment across distributions.
	\item Rapid adaptation to new subject's data during testing.
	\item Mitigation of privacy leakage risks caused by transmitting the statistical parameters, effectively defending against attacks that attempt to reverse-engineer raw data through BN parameters.
\end{enumerate}

\subsection{FAT} \label{sec:fat}

AT enhances model robustness against adversarial attacks by incorporating adversarial examples during training. It solves the following min-max optimization problem:
\begin{align}
 \min_{\boldsymbol\theta}\mathbb{E}_{(\mathbf{X},y)\sim \mathcal{D}}\left[\max_{\Vert\boldsymbol\delta\Vert_p\leq\epsilon }\mathcal{L}\left(\boldsymbol\theta,\mathbf{X}+\boldsymbol\delta,y\right)\right],
\end{align}
where $\boldsymbol\theta$ denotes model parameters, $\mathcal{D}$ the training set, $\mathcal{L}$ the loss function, and $\boldsymbol{\delta}$ the adversarial perturbation constrained within an $\epsilon$-bounded $\ell_p$-norm ball. The input $\mathbf{X} + \boldsymbol{\delta}$ is the adversarial counterpart of $\mathbf{X}_{adv}$. The inner maximization generates worst-case perturbations, and the outer minimization optimizes model parameters against them.

FAT extends classical AT to the FL framework through three key steps: 1) generating adversarial examples from local client data during training; 2) performing local AT using the adversarial samples; and, 3) aggregating the client model parameters at the central server. It imposes $\ell_\infty$-norm constraints on adversarial perturbations as $\|\boldsymbol\delta\|_\infty \leq \epsilon$. 

To address the limited computational resources on clients in federated systems, FAT employs the Fast Gradient Sign Method (FGSM)~\cite{Goodfellow2015Explaining} for efficient perturbation generation:
\begin{align} \label{eq:fgsm}
	\boldsymbol\delta = \alpha\cdot \text{sign}\left(\nabla_{\mathbf{X}} \mathcal{L}\left(\boldsymbol{\theta}, \mathbf{X}, y\right)\right),
\end{align}
where $\alpha\leq\epsilon$ controls the perturbation magnitude, under the $\ell_\infty$-norm constraint. This single-step computation enables efficient gradient acquisition through backpropagation, making FAT particularly suitable for distributed learning scenarios with resource-constrained clients.

\subsection{AWP}

FAT enhances model robustness by introducing perturbations in the input space to smooth the loss landscape. AWP~\cite{Wu2020Adversarial} is further introduced into the client-side training of FL, to regularize the flatness of the weight loss landscape through parameter space perturbations, and hence to strengthen the adversarial robustness.

AWP solves a min-max optimization problem:
\begin{align}
	\begin{split}
		&\min_{\boldsymbol\theta} \{\mathcal{L}(\boldsymbol\theta)+\max_{\Vert \boldsymbol\nu_l\Vert \leq \xi \Vert \boldsymbol\theta_l\Vert}\left[\mathcal{L}\left(\boldsymbol\theta+\boldsymbol\nu\right)-\mathcal{L}(\boldsymbol\theta)\right]\} \\
		=&\min_{\boldsymbol\theta}\max_{\Vert \boldsymbol\nu_l\Vert \leq \xi \Vert \boldsymbol\theta_l\Vert}\mathcal{L}(\boldsymbol\theta+\boldsymbol\nu),
	\end{split}
\end{align}
where $\mathcal{L}(\boldsymbol\theta)$ (a simplified notation for $\mathcal{L}(\boldsymbol\theta, \mathbf{X}, y)$) represents the original training loss, $\mathcal{L}(\boldsymbol\theta+\boldsymbol\nu)-\mathcal{L}(\boldsymbol\theta)$ characterizes the flatness of the weight loss landscape, $\boldsymbol\nu$ is the weight perturbation, and $\Vert \boldsymbol\nu_l\Vert \leq \xi \Vert \boldsymbol\theta_l\Vert$ is the constraint on the perturbation magnitude (proportional to layer-wise weights $\boldsymbol\theta_l$), with $\xi$ controlling the perturbation scale.

To accommodate the limited computational capacity of FL clients, we adopt a simplified one-step approximation to compute $\boldsymbol\nu$:
\begin{align}\label{eq:nu}
	\boldsymbol\nu \leftarrow \xi \frac{\nabla_{\boldsymbol\nu}\mathcal{L}\left(\boldsymbol\theta+\boldsymbol\nu\right)}{\Vert \nabla_{\boldsymbol\nu}\mathcal{L}\left(\boldsymbol\theta+\boldsymbol\nu\right)\Vert }\Vert \boldsymbol\theta\Vert,
\end{align}
and then update the model weights by:
\begin{align}\label{eq:theta}
	\boldsymbol\theta \leftarrow (\boldsymbol\theta+\boldsymbol\nu)-\eta\nabla_{\boldsymbol\theta+\boldsymbol\nu}\mathcal{L}(\boldsymbol\theta+\boldsymbol\nu)-\boldsymbol\nu,
\end{align}
where $\eta$ is the learning rate, and the parameters are reset to their original values after each perturbation step.

\section{Experiment Settings} \label{sec:settings}

This section introduces our experiment settings, including datasets, preprocessing, threat models, baseline algorithms, evaluation metrics, and hyperparameters.

\subsection{Datasets and Preprocessing}

Three MI and two ERP datasets, summarized in Table~\ref{tab:dataset}, were used in our experiments.

\begin{table}[htbp] \centering \setlength{\tabcolsep}{1.2mm}
	\caption{Summary of the five datasets.}    \label{tab:dataset}
	\begin{tabular}{c|ccccc}
		\toprule
		Dataset &\begin{tabular}[c]{@{}c@{}}\# Subjects\end{tabular} & \begin{tabular}[c]{@{}c@{}}\# Classes\end{tabular} & \begin{tabular}[c]{@{}c@{}}\# Channels\end{tabular} & \begin{tabular}[c]{@{}c@{}}\# Trials\\ per Subjects\end{tabular} & \begin{tabular}[c]{@{}c@{}}Class-\\ Imbalance\end{tabular} \\ \midrule
		MI1~\cite{Yi2014Evaluation}     & 10          & 6          & 60          & 480       & \XSolidBrush              \\
		MI2~\cite{Tangermann2012Review}     & 9           & 4          & 22          & 576       & \XSolidBrush              \\
		MI3~\cite{Steyrl2016Random}     & 14          & 2          & 15          & 160       & \XSolidBrush              \\
		ERP1~\cite{Arico2014Influence}    & 10          & 2          & 16          & 1728      & \Checkmark             \\
		ERP2~\cite{Riccio2013Attention}    & 8           & 2          & 8           & 4200      & \Checkmark              \\ \bottomrule
	\end{tabular}
\end{table}

MI1~\cite{Yi2014Evaluation} includes six classes (left hand, right hand, both feet, both hands, left hand combined with right foot, and right hand combined with left foot), each with 80 trials. 60-channel EEG signals from 10 healthy subjects were recorded at 200 Hz. We resampled the signals to 250 Hz, extracted data from [0,4] seconds following task stimulation, and applied a [8–30] Hz band-pass filter.

MI2~\cite{Tangermann2012Review} was originally released as BCI Competition IV dataset 2a. It includes four classes (left hand, right hand, both feet, and tongue), each with 144 trials. 22-channel EEG signals from 9 healthy subjects were recorded at 250 Hz. We extracted data from [0,4] seconds following task stimulation and applied a [8–30] Hz band-pass filter.

MI3~\cite{Steyrl2016Random} includes two classes (left hand versus right hand), each with 80 trials. 15-channel EEG signals from 14 subjects were sampled at 512 Hz. We downsampled the signals to 250 Hz, extracted data from [0,5] seconds following task stimulation, and applied a [8–30] Hz band-pass filter.

ERP1~\cite{Arico2014Influence} includes two classes (non-target and target) with a 5:1 ratio, with 1,440 non-target trials and 288 target trials. 16-channel EEG signals from 10 healthy subjects were recorded at 256 Hz. We downsampled the signals to 250 Hz, extracted data from [0,0.8] seconds following event stimulation, and applied a [0.1,20] Hz band-pass filter.

ERP2~\cite{Riccio2013Attention} includes two classes (non-target and target) with a 5:1 ratio, with 3500 non-target trials and 700 target trials. 8-channel EEG signals from 8 amyotrophic lateral sclerosis patients were recorded at 256 Hz. We downsampled the signals to 250 Hz, extracted data from [0,1] seconds following event stimulation, and applied a [0.1,30] Hz band-pass filter.

\subsection{Threat Models}

We considered three white-box adversarial attacks:
\begin{enumerate}
\item FGSM~\cite{Goodfellow2015Explaining}, which computes the gradient of the loss with respect to the input and applies a small perturbation along the direction of the sign of the loss gradient:
	\begin{align}
		\mathbf{X}_{adv} = \mathbf{X} + \alpha \cdot \text{sign}(\nabla_{\mathbf{X}} \mathcal{L}(\boldsymbol{\theta}, \mathbf{X}, y)).
	\end{align}

\item Projected gradient descent (PGD)~\cite{Madry2018Towards}, which iteratively improves FGSM. Starting from an initial perturbation (often randomized) applied to the input $\mathbf{X}$, the adversarial example is updated iteratively:
	\begin{align}
		\begin{split}
			\mathbf{X}^{t+1}_{adv} = \text{Proj}_{\mathbf{X},\epsilon} \left( \mathbf{X}^t_{adv} + \alpha \cdot \text{sign}(\nabla_{\mathbf{X}} \mathcal{L}(\boldsymbol{\theta}, \mathbf{X}^t_{adv}, y)) \right),
		\end{split}
	\end{align}
where $\mathbf{X}^t_{adv}$ denotes the adversarial example in iteration $t$, $\alpha$ controls the step size, and $\text{Proj}_{\mathbf{X},\epsilon}(\cdot)$ projects the perturbation within the $\ell_\infty$-norm ball of radius $\epsilon$ centered at $\mathbf{X}$. This multi-step exploration results in stronger attacks than the single-step FGSM.

	\item AutoAttack (AA)~\cite{Croce2020Reliable}, a parameter-free ensemble evaluation method combining four complementary attack strategies: Auto-PGD with cross-entropy loss, Auto-PGD with difference of logits ratio loss, Fast Adaptive Boundary attack, and Square attack~\cite{Andriushchenko2020Square}. This combination ensures robustness across diverse scenarios by leveraging both white-box gradient information and black-box query strategies.
\end{enumerate}
and two black-box adversarial attacks:
\begin{enumerate}
	\item RayS\cite{Chen2020RayS}, an efficient hard-label attack that replaces continuous optimization with discrete search over $\ell_\infty$-norm vertices $\boldsymbol{\delta} \in \{-1,1\}^d$, where $d$ denotes the input dimensionality. It seeks minimal perturbation radius via greedy dimension-wise adjustments with early direction pruning to reduce the number of queries, and hierarchical coarse-to-fine search to leverage the spatial correlations.

	\item Square attack~\cite{Andriushchenko2020Square}, a score-based black-box adversarial attack method applicable to the $\ell_{\infty}$ norm. It employs a randomized search strategy, selecting localized square-shaped updates at random positions, ensuring that the perturbation at each iteration is approximately situated at the boundary of the feasible set. Square attack demonstrates high query efficiency and attack accuracy, particularly in untargeted attacks.
\end{enumerate}

All attacks adopt the $\ell_{\infty}$ untargeted setting, aiming to induce misclassification rather than targeting specific classes. Given the characteristics of EEG signals, we set the perturbation magnitude as $\epsilon$ times the signal standard deviation~\cite{Chen2024Alignment}. For the three white-box attacks, we selected $\epsilon \in \{0.01, 0.03, 0.05\}$. For the two black-box attacks, which have limited access to model information, we chose $\epsilon \in \{0.01, 0.05, 0.1\}$.

\subsection{Baseline Algorithms}

We compared SAFE with 14 state-of-the-art CT and FL approaches, covering a wide range of methods designed to enhance adversarial robustness and generalization. All approaches employed EEGNet~\cite{Lawhern2018EEGNet} as the backbone architecture.

CT approaches merge data from $K$ subjects for joint training, without considering any privacy protection. The seven CT baselines include:
\begin{enumerate}
	\item Normal Training (NT), which applies standard empirical risk minimization to the pooled training data from all subjects.
	\item AT, which enhances adversarial robustness by augmenting the training data with FGSM-generated adversarial examples.
	\item Five domain generalization approaches, including DANN~\cite{Ganin2016Domain} based on domain adversarial learning, IRM~\cite{Arjovsky2019Invariant} based on invariant risk minimization, GroupDRO~\cite{Sagawa2020Distributionally} based on distributionally robust optimization, MLDG~\cite{Li2018Learning} based on meta-learning, and RSC~\cite{Huang2020Self} based on gradient manipulation.
\end{enumerate}

FL treats each training subject as a separate client for privacy protection. The seven FL baselines include:
\begin{enumerate}
	\item FedAvg~\cite{McMahan2017Communication}, the most widely used FL algorithm, where each client performs multiple stochastic gradient descent update steps in a single communication round to balance communication cost and accuracy.
	\item FAL~\cite{Li2023Federated}, which uses PGD-generated adversarial samples for adversarial learning in a federated environment, enhancing robustness against adversarial attacks.
	\item Three approaches focusing on mitigating client data drift, including FedProx~\cite{Li2020Federated}, which introduces a proximal term in client objectives to reduce the discrepancy between local and global models; SCAFFOLD~\cite{Karimireddy2020Scaffold}, which employs control variables to correct client drift during local updates; and MOON~\cite{Li2021Model}, which leverages model representation similarity during optimization.
	\item Two methods focusing on improving model generalization, including FedFA~\cite{Zhou2023FedFA}, which performs federated feature augmentation to enhance generalization, and GA~\cite{Zhang2023Federated}, which incorporates domain generalization through fairness objectives and requires all clients to participate in every communication round.
\end{enumerate}

\subsection{Evaluation Metrics}

We employed balanced classification accuracy (BCA) to assess the model performance on both benign and adversarial samples, because the conventional raw classification accuracy becomes unreliable for highly imbalanced datasets like ERP1 and ERP2. 

The BCA is computed as:
\begin{align}
	\text{BCA} = \frac{1}{L}\frac{1}{N_l}\sum_{l=1}^L\sum_{i=1}^{N_l}\mathbf{1}(\hat{y}_i=y_i),
\end{align}
where $L$ denotes the total number of classes, $N_l$ the number of test samples in class $l$, $\hat{y}_i$ the predicted label of the $i$-th test sample, $y_i$ the true label, and $\mathbf{1}(\cdot)$ the indicator function.

\subsection{Additional Settings and Hyperparameters}

Leave-one-subject-out cross-validation was adopted. Specifically, each subject in the corresponding dataset was treated as an unknown test subject once, while the remaining subjects served as training data. In the CT setting, data from all training subjects were aggregated for model training. In the FL setting, each client represented one training subject, i.e., the number of clients equaled the total number of subjects minus one. Five repeats with different random seeds were performed, and the average results are reported.

Euclidean alignment (EA)~\cite{He2020Transfer} was applied individually to each subject for all approaches. 

To mitigate class imbalance in ERP1 and ERP2, we implemented a mix-based data augmentation strategy~\cite{Chen2025Data} for all approaches within each subject. Specifically, two EEG trials $\mathbf{X}_1$ and $\mathbf{X}_2$ were randomly selected, and their arithmetic mean was computed as the augmented sample $\mathbf{X}'$:
\begin{align}
	\mathbf{X}' = \frac{1}{2}(\mathbf{X}_1 + \mathbf{X}_2).
\end{align}
$\mathbf{X}'$ was labeled as the target class if at least one of $\mathbf{X}_1$ or $\mathbf{X}_2$ belonged to the target class, and as non-target class otherwise.

Stochastic gradient descent optimizer was employed with a learning rate of 0.005, weight decay of 1e-4, and momentum of 0.9. For CT, models were trained for 100 epochs with varying batch sizes: 64 for MI, and 256/512 for ERP1/ERP2, respectively. In FL, we conducted 100 global communication rounds where half of the clients were randomly selected per round to reduce the communication overhead, and each client performed two local training epochs with reduced batch sizes (32 for MI, and 128/256 for ERP1/ERP2) to accommodate limited client-side samples. During testing phases for both benign and adversarial samples, a fixed batch size of 8 was used across all experiments to ensure consistency.

For SAFE, the perturbation magnitude $\alpha$ in FAT was set to $0.03$ times the signal's standard deviation. The perturbation scale for AWP was set to $\xi = 0.01$. For AT, FGSM generated adversarial examples with a perturbation magnitude of $0.03$ times the signal's standard deviation. Hyperparameters of other baselines were set according to the recommendations in their original publications and further fine-tuned slightly.

\section{Experimental Results} \label{sec:results}

This section presents experimental results to demonstrate the effectiveness of SAFE. 

\subsection{BCAs on Benign Samples}

Tables~\ref{tab:benign_result_mi1}-\ref{tab:benign_result_erp2} present the cross-subject BCAs for benign samples on the five datasets. The best performance in each column is marked in bold, and the second-best underlined. 

\begin{table*}[htbp] \centering \setlength{\tabcolsep}{3mm}
	\caption{Cross-subject BCAs (\%) for benign samples on MI1. The best accuracy in each column is marked in bold, and the second best underlined.}.    \label{tab:benign_result_mi1}
	\begin{tabular}{c|c|cccccccccc|c}
		\toprule
		Setting & Approach & S1    & S2    & S3    & S4    & S5    & S6    & S7    & S8    & S9    & S10   & Avg.  \\ \midrule
		\multirow{7}{*}{CT} & NT     & 35.17 & 37.38 & 32.13 & 24.58 & {\ul34.79} & \textbf{52.67} & \textbf{42.67} & {\ul46.92} & \textbf{58.33} & {\ul23.83} & {\ul38.85} \\
		& AT       & 23.04 & 30.75 & 25.92 & 20.25 & 26.71 & 36.90 & 36.04 & 33.83 & 41.75 & 17.50 & 29.27 \\
		& DANN     & 26.08 & 33.42 & 26.92 & 24.33 & 29.54 & 45.05 & 38.00 & 40.71 & 47.96 & 23.17 & 33.52 \\
		& IRM      & 34.04 & 36.58 & 29.08 & 23.83 & 30.42 & 47.67 & 41.04 & 39.42 & 49.67 & 22.38 & 35.41 \\
		& GroupDro & {\ul37.25} & 34.58 & 31.29 & 24.96 & 30.46 & 50.00 & 38.42 & 44.21 & 53.58 & 23.50 & 36.83 \\
		& MLDG     & 26.08 & 28.04 & 24.33 & 19.13 & 25.08 & 26.52 & 37.00 & 37.25 & 44.38 & 20.25 & 28.81 \\
		& RSC      & 32.25 & 35.04 & 32.00 & 22.17 & 29.54 & 49.71 & {\ul42.29} & 39.42 & 54.00 & 22.58 & 35.90 \\ \midrule
		\multirow{7}{*}{FL} & FedAvg & 29.71 & {\ul39.46} & 27.75 & {\ul25.33} & 26.21 & 51.57 & 37.67 & 46.33 & 51.71 & 19.54 & 35.53 \\
		& FAL      & 19.25 & 31.63 & 25.00 & 20.38 & 20.88 & 37.67 & 32.50 & 27.29 & 28.58 & 16.17 & 25.93 \\
		& FedProx  & 30.54 & 36.92 & 29.92 & 25.04 & 25.25 & 47.29 & 37.42 & 41.04 & 53.71 & 17.88 & 34.50 \\
		& SCAFFOLD & 17.71 & 16.21 & 17.42 & 15.88 & 18.04 & 17.76 & 18.08 & 18.00 & 20.63 & 16.67 & 17.64 \\
		& MOON     & 33.96 & 38.21 & {\ul32.75} & \textbf{25.63} & 33.71 & 50.00 & 41.21 & 45.04 & 53.00 & 20.46 & 37.40 \\
		& FedFA    & 29.88 & 31.58 & 26.79 & 24.00 & 20.79 & 42.81 & 27.63 & 31.25 & 38.87 & 18.04 & 29.16 \\
		& GA       & 30.71 & 35.38 & 27.54 & 24.33 & 26.42 & 46.05 & 31.21 & 40.58 & 50.33 & 19.83 & 33.24 \\
		& SAFE     & \textbf{39.42} & \textbf{45.38} & \textbf{36.04} & 23.83 & \textbf{34.83} & {\ul52.62} & 40.63 & \textbf{58.25} & {\ul57.75} & \textbf{29.88} & \textbf{41.86} \\ \bottomrule
	\end{tabular}
\end{table*}

\begin{table*}[htbp] \centering \setlength{\tabcolsep}{3.5mm}
	\caption{Cross-subject BCAs (\%) for benign samples on MI2. The best accuracy in each column is marked in bold, and the second best underlined.}.    \label{tab:benign_result_mi2}
	\begin{tabular}{c|c|ccccccccc|c}
		\toprule
		Setting & Approach             & S1    & S2    & S3    & S4    & S5    & S6    & S7    & S8    & S9    & Avg.  \\ \midrule
		\multirow{7}{*}{CT} & NT       & \textbf{69.17} & 32.78 & 68.92 & 41.25 & 34.58 & \textbf{41.81} & 42.40 & 65.73 & 60.97 & 50.84 \\
		& AT       & 62.19 & 32.26 & \textbf{73.44} & 40.00 & 33.58 & 36.84 & {\ul51.77} & \textbf{66.18} & {\ul64.13} & {\ul51.15} \\
		& DANN     & 57.36 & 31.28 & 55.56 & 37.88 & 31.11 & 38.99 & 38.78 & 53.44 & 55.83 & 44.47 \\
		& IRM      & 62.15 & 30.80 & 62.26 & 40.69 & 31.81 & 38.54 & 36.98 & 55.10 & 54.48 & 45.87 \\
		& GroupDro & 62.50 & {\ul33.13} & 59.38 & {\ul41.49} & {\ul35.52} & {\ul39.86} & 33.92 & 54.06 & 51.42 & 45.70 \\
		& MLDG     & 53.68 & 28.19 & 60.97 & 27.29 & 32.53 & 28.99 & 40.69 & 48.13 & 53.37 & 41.54 \\
		& RSC      & {\ul66.88} & 30.94 & 71.08 & 38.85 & 32.74 & 37.33 & 40.31 & 60.94 & 58.19 & 48.58 \\ \midrule
		\multirow{7}{*}{FL} & FedAvg   & 60.59 & 29.93 & 54.97 & 37.40 & 31.15 & 39.06 & 36.60 & 46.94 & 52.15 & 43.20 \\
		 & FAL      & 54.90 & 30.56 & 55.17 & 36.88 & 31.35 & 37.12 & 36.04 & 52.57 & 58.96 & 43.73 \\
		& FedProx  & 53.33 & 28.61 & 54.24 & 35.42 & 29.38 & 37.40 & 38.26 & 45.00 & 42.60 & 40.47 \\
		& SCAFFOLD & 41.22 & 26.81 & 39.17 & 29.27 & 27.60 & 29.06 & 27.15 & 31.91 & 40.10 & 32.48 \\
		& MOON     & 64.24 & 30.80 & 68.26 & 37.67 & 34.31 & 38.47 & 43.65 & 58.40 & 54.13 & 47.77 \\
		& FedFA    & 59.55 & \textbf{33.61} & 56.32 & 39.62 & 30.90 & 38.06 & 40.03 & 46.88 & 53.65 & 44.29 \\
		& GA       & 45.59 & 27.01 & 48.19 & 36.32 & 31.81 & 39.76 & 32.15 & 32.50 & 44.76 & 37.57 \\
		& SAFE     & 65.80 & 30.63 & {\ul72.53} & \textbf{41.91} & \textbf{36.67} & 38.06 & \textbf{62.99} & {\ul66.08} & \textbf{66.56} & \textbf{53.47} \\ \bottomrule
	\end{tabular}
\end{table*}

\begin{table*}[htbp] \centering \setlength{\tabcolsep}{1.8mm}
	\caption{Cross-subject BCAs (\%) for benign samples on MI3. The best accuracy in each column is marked in bold, and the second best underlined.}.    \label{tab:benign_result_mi3}
	\begin{tabular}{c|c|cccccccccccccc|c}
		\toprule
		Setting & Approach & S1    & S2    & S3    & S4    & S5    & S6    & S7    & S8    & S9    & S10   & S11   & S12   & S13   & S14   & Avg.  \\ \midrule
		\multirow{7}{*}{CT} & NT     & 68.25 & 84.75 & 90.13 & \textbf{87.63} & 76.75 & 70.25 & {\ul94.00} & {\ul72.50} & 93.50 & {\ul68.25} & \textbf{85.25} & 79.88 & {\ul63.00} & 50.38 & {\ul77.46} \\
		& AT       & 62.50 & 83.50 & 81.00 & 77.88 & {\ul80.63} & \textbf{76.38} & 93.13 & 66.63 & 92.00 & \textbf{68.50} & 81.50 & 76.00 & 58.88 & 50.13 & 74.90 \\
		& DANN     & 64.63 & 82.63 & {\ul93.88} & 82.25 & 65.63 & 66.13 & 91.63 & 71.00 & 93.75 & 63.00 & 74.75 & 71.13 & \textbf{63.13} & \textbf{54.88} & 74.17 \\
		& IRM      & 65.13 & 83.38 & 87.63 & 83.50 & 74.50 & 72.50 & 91.13 & 69.50 & 93.50 & 68.00 & 80.13 & 79.63 & 60.13 & 50.25 & 75.63 \\
		& GroupDro & 64.38 & 82.38 & 71.00 & 81.50 & 71.00 & 68.25 & 89.50 & 60.50 & 92.88 & 65.75 & 82.50 & 70.88 & 59.38 & 48.13 & 72.00 \\
		& MLDG     & 58.38 & 61.63 & 64.75 & 63.63 & 66.88 & 55.50 & 77.50 & 50.88 & 91.38 & 57.50 & 71.13 & 60.00 & 51.63 & 50.25 & 62.93 \\
		& RSC      & {\ul70.13} & {\ul85.25} & 91.00 & 84.38 & 76.88 & 73.75 & \textbf{94.50} & 67.88 & 93.50 & 67.12 & {\ul84.25} & \textbf{83.63} & 61.50 & 48.25 & 77.29 \\ \midrule
		\multirow{7}{*}{FL} & FedAvg & 58.50 & 83.88 & 55.75 & 83.25 & \textbf{82.13} & 69.50 & 91.00 & 52.88 & 95.13 & 64.75 & 79.25 & 66.25 & 58.88 & 49.75 & 70.78 \\
		& FAL      & 55.25 & 74.63 & 57.00 & 65.25 & 76.25 & 62.75 & 84.38 & 55.88 & 93.63 & 65.13 & 75.75 & 61.38 & 57.25 & 49.25 & 66.70 \\
		& FedProx  & 57.25 & 82.38 & 56.38 & 73.25 & {\ul80.63} & 65.00 & 86.88 & 51.63 & \textbf{95.50} & 65.25 & 80.13 & 68.00 & 59.88 & 49.88 & 69.43 \\
		& SCAFFOLD & 67.13 & 82.13 & 61.50 & 74.25 & 68.88 & 70.38 & 90.00 & 62.38 & 94.38 & 66.63 & 74.50 & 56.00 & 59.63 & {\ul52.25} & 70.00 \\
		& MOON     & 61.38 & \textbf{85.63} & 66.75 & {\ul85.50} & 73.13 & 71.88 & 93.63 & 54.63 & {\ul95.38} & 66.63 & 83.88 & 81.50 & 60.63 & 49.38 & 73.56 \\
		& FedFA    & 60.75 & 79.88 & 59.13 & 80.13 & 79.00 & 68.13 & 87.38 & 59.13 & 94.88 & 65.38 & 81.00 & 71.38 & 60.00 & 48.63 & 71.05 \\
		& GA       & 54.25 & 81.38 & 52.00 & 79.38 & 74.13 & 62.38 & 87.13 & 52.50 & 95.13 & 62.63 & 74.63 & 57.88 & 58.25 & 50.50 & 67.29 \\
		& SAFE     & \textbf{71.50} & 82.38 & \textbf{94.75} & 81.38 & {\ul80.63} & {\ul75.50} & 91.25 & \textbf{80.75} & 93.38 & \textbf{68.50} & 82.13 & {\ul82.63} & 60.50 & 47.50 & \textbf{78.05} \\ \bottomrule
	\end{tabular}
\end{table*}

\begin{table*}[htbp] \centering \setlength{\tabcolsep}{3mm}
	\caption{Cross-subject BCAs (\%) for benign samples on ERP1. The best accuracy in each column is marked in bold, and the second best underlined.}.    \label{tab:benign_result_erp1}
	\begin{tabular}{c|c|cccccccccc|c}
		\toprule
		Setting & Approach & S1    & S2    & S3    & S4    & S5    & S6    & S7    & S8    & S9    & S10   & Avg.  \\ \midrule
		\multirow{7}{*}{CT} & NT     & {\ul74.56} & \textbf{86.97} & 77.88 & 81.99 & 83.35 & 79.09 & {\ul79.77} & 73.78 & 85.97 & 84.35 & 80.77 \\
		& AT       & 74.47 & 82.08 & 74.45 & 82.47 & 81.72 & {\ul80.07} & 79.63 & 72.34 & 83.71 & 81.01 & 79.19 \\
		& DANN     & 73.85 & {\ul86.24} & 77.69 & 81.81 & 83.43 & 78.76 & 77.10 & 72.34 & 84.54 & 84.24 & 80.00 \\
		& IRM      & 73.64 & 85.81 & 78.01 & \textbf{83.62} & 85.44 & 78.26 & 78.77 & 73.96 & 86.23 & 85.12 & 80.88 \\
		& GroupDro & 69.26 & 85.03 & 76.68 & 83.19 & 84.19 & 78.28 & 77.14 & 73.54 & 83.69 & 84.10 & 79.51 \\
		& MLDG     & 64.25 & 67.43 & 63.08 & 65.33 & 66.50 & 68.54 & 60.78 & 58.44 & 59.86 & 63.20 & 63.74 \\
		& RSC      & 73.41 & 85.95 & \textbf{78.46} & {\ul83.55} & 86.34 & 79.24 & 78.38 & {\ul74.64} & 86.17 & {\ul84.77} & {\ul81.09} \\ \midrule
		\multirow{7}{*}{FL} & FedAvg & 70.54 & 79.15 & 75.15 & 80.93 & 87.28 & 75.51 & 66.63 & 70.19 & 80.24 & \textbf{85.92} & 77.15 \\
		& FAL      & 73.94 & 82.35 & 76.30 & 81.74 & \textbf{88.65} & 79.58 & 70.10 & 72.89 & 83.74 & 83.49 & 79.28 \\
		& FedProx  & 71.99 & 82.06 & 76.77 & 81.30 & {\ul87.95} & 78.94 & 71.49 & 72.79 & 84.34 & \textbf{85.92} & 79.36 \\
		& SCAFFOLD & 63.93 & 79.69 & 74.52 & 76.36 & 82.22 & 77.96 & 76.17 & 70.37 & 80.70 & 78.91 & 76.08 \\
		& MOON     & 72.23 & 83.49 & 76.40 & 81.48 & 87.30 & 78.79 & 74.36 & 74.12 & 84.98 & 82.94 & 79.61 \\
		& FedFA    & 73.42 & 84.16 & {\ul78.06} & 81.90 & 87.77 & 78.41 & 74.13 & 74.11 & \textbf{86.59} & 84.22 & 80.28 \\
		& GA       & 68.49 & 78.81 & 75.63 & 82.32 & 84.67 & 74.42 & 67.53 & 70.87 & 80.40 & 84.46 & 76.76 \\
		& SAFE     & \textbf{75.71} & 84.28 & 77.42 & 83.05 & 87.94 & \textbf{81.35} & \textbf{80.09} & \textbf{75.63} & {\ul86.58} & 84.13 & \textbf{81.62} \\ \bottomrule
	\end{tabular}
\end{table*}

\begin{table*}[htbp] \centering \setlength{\tabcolsep}{4.5mm}
	\caption{Cross-subject BCAs (\%) for benign samples on ERP2. The best accuracy in each column is marked in bold, and the second best underlined.}    \label{tab:benign_result_erp2}
	\begin{tabular}{c|c|cccccccc|c}
		\toprule
		Setting & Approach             & S1    & S2    & S3    & S4    & S5    & S6    & S7    & S8    & Avg.  \\ \midrule
		\multirow{7}{*}{CT} & NT       & 71.46 & 67.54 & 73.04 & 68.71 & 71.97 & {\ul70.44} & 71.95 & 71.60 & 70.84 \\
		& AT       & 70.03 & 66.35 & 70.46 & 67.57 & 70.11 & 69.40 & \textbf{72.30} & 70.23 & 69.55 \\
		& DANN     & 70.83 & 66.70 & 72.90 & 67.75 & 70.21 & 68.77 & 70.51 & 70.70 & 69.80 \\
		& IRM      & 71.95 & 68.55 & 74.07 & 69.02 & \textbf{73.07} & 70.43 & {\ul72.07} & 70.41 & {\ul71.20} \\
		& GroupDro & 71.25 & \textbf{69.20} & 74.41 & 68.83 & 71.89 & 69.24 & 71.53 & 68.09 & 70.55 \\
		& MLDG     & 54.64 & 50.43 & 53.25 & 52.20 & 51.81 & 51.80 & 51.85 & 50.09 & 52.01 \\
		& RSC      & {\ul71.98} & 67.68 & 74.03 & 69.30 & {\ul72.45} & \textbf{70.65} & 71.72 & 71.05 & 71.11 \\ \midrule
		\multirow{7}{*}{FL} & FedAvg   & 70.74 & 66.52 & {\ul75.04} & 70.26 & 72.08 & 68.62 & 70.23 & 72.77 & 70.78 \\
		& FAL      & 69.83 & 66.05 & 74.44 & \textbf{70.81} & 70.95 & 68.66 & 71.25 & 72.40 & 70.55 \\
		& FedProx  & 71.13 & 65.52 & 74.99 & {\ul70.28} & 72.25 & 68.12 & 70.91 & 71.31 & 70.57 \\
		& SCAFFOLD & 66.27 & 61.76 & 66.73 & 65.39 & 65.07 & 65.61 & 66.05 & 66.75 & 65.45 \\
		& MOON     & 70.79 & 65.66 & 74.52 & 70.07 & 71.93 & 68.25 & 69.27 & 71.28 & 70.22 \\
		& FedFA    & \textbf{72.17} & 66.23 & \textbf{75.05} & 69.81 & 71.88 & 69.34 & 70.09 & {\ul73.37} & 70.99 \\
		& GA       & 67.89 & 64.72 & 73.52 & 69.44 & 71.14 & 69.06 & 70.20 & \textbf{74.14} & 70.01 \\
		& SAFE     & 70.83 & {\ul68.60} & 74.73 & \textbf{70.81} & 72.03 & 70.01 & 71.59 & 72.99 & \textbf{71.45} \\ \bottomrule
	\end{tabular}
\end{table*}

Observe that:
\begin{enumerate}
	\item In the CT setting, NT achieved the best average accuracy, outperforming domain generalization methods, indicating the limitations of existing domain generalization approaches in addressing cross-subject EEG distribution discrepancies.
	\item Since the primary goal of both AT and FAL is to enhance the model robustness to adversarial samples instead of the decoding accuracy on the benign samples, employing either AT or FAL alone led to decreased classification accuracy on the benign samples.
	\item All FL methods, except SAFE, underperformed CT, indicating that client data heterogeneity caused performance degradation and exposed the conflict between privacy protection and decoding accuracy.
	\item SAFE achieved the best average performance across all five datasets, achieving both accurate decoding and privacy protection.
\end{enumerate}

\subsection{BCAs on Adversarial Samples}

\begin{figure}[htbp] 	\centering
	{\includegraphics[width=\linewidth]{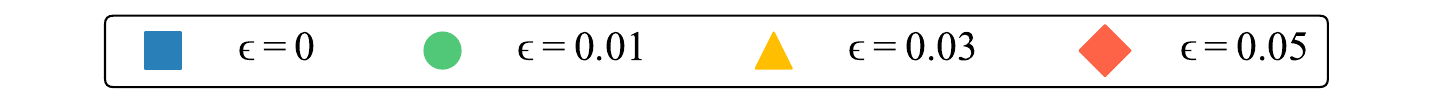}}
	\subfigure[MI1]{\label{fig:mi1_white}\includegraphics[width=\linewidth]{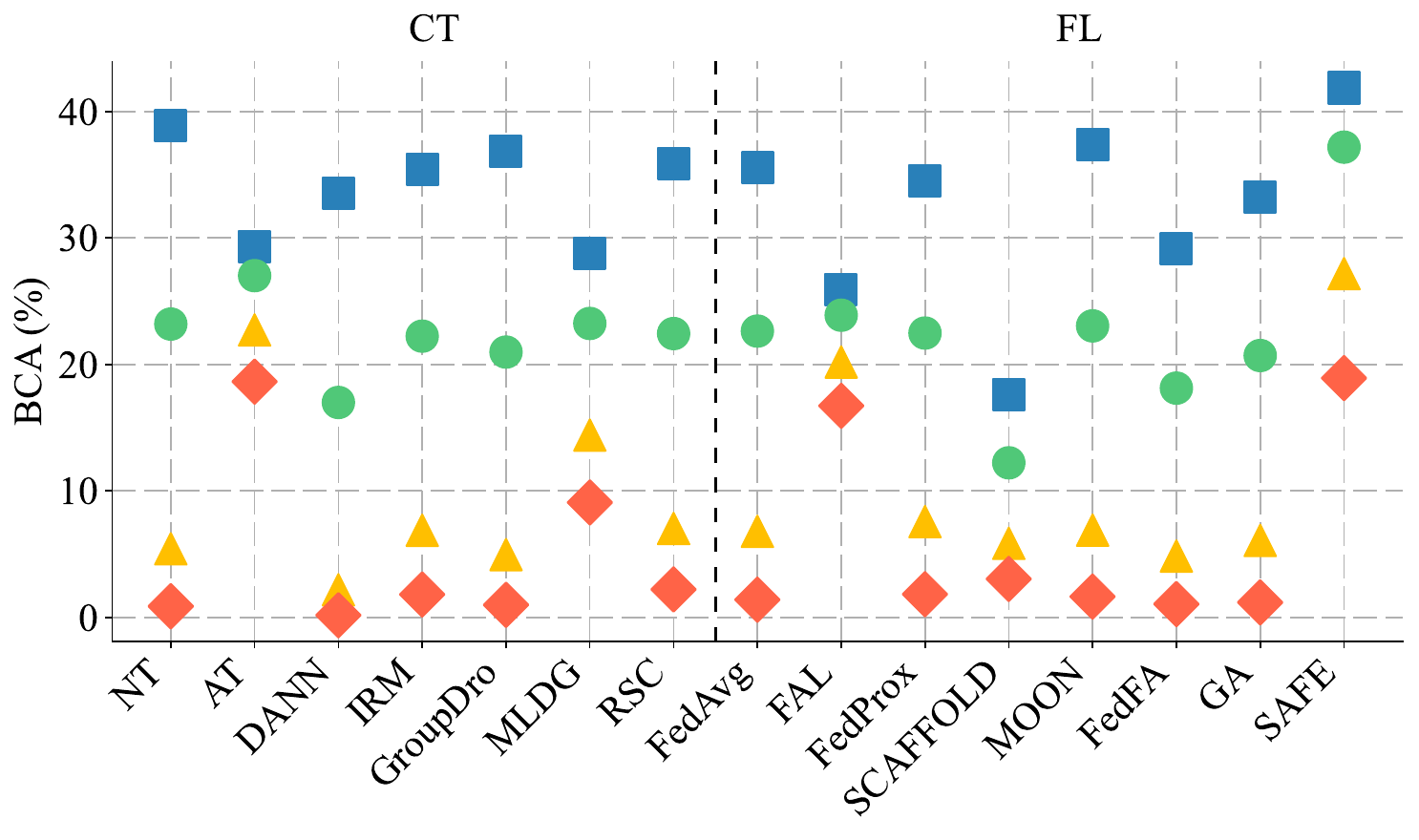}}
	\subfigure[MI2]{\label{fig:mi2_white}\includegraphics[width=\linewidth]{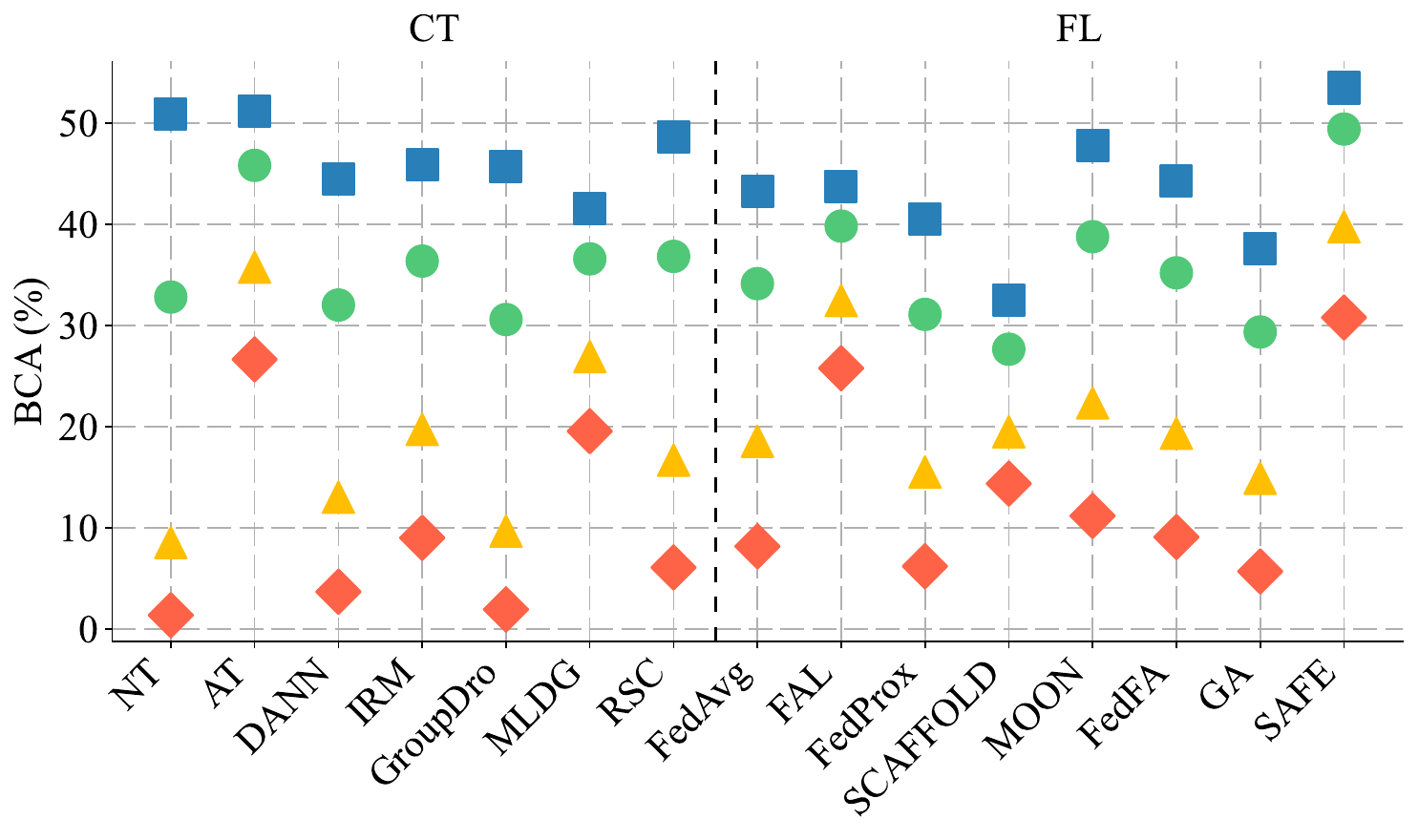}}
	\subfigure[MI3]{\label{fig:mi3_white}\includegraphics[width=\linewidth]{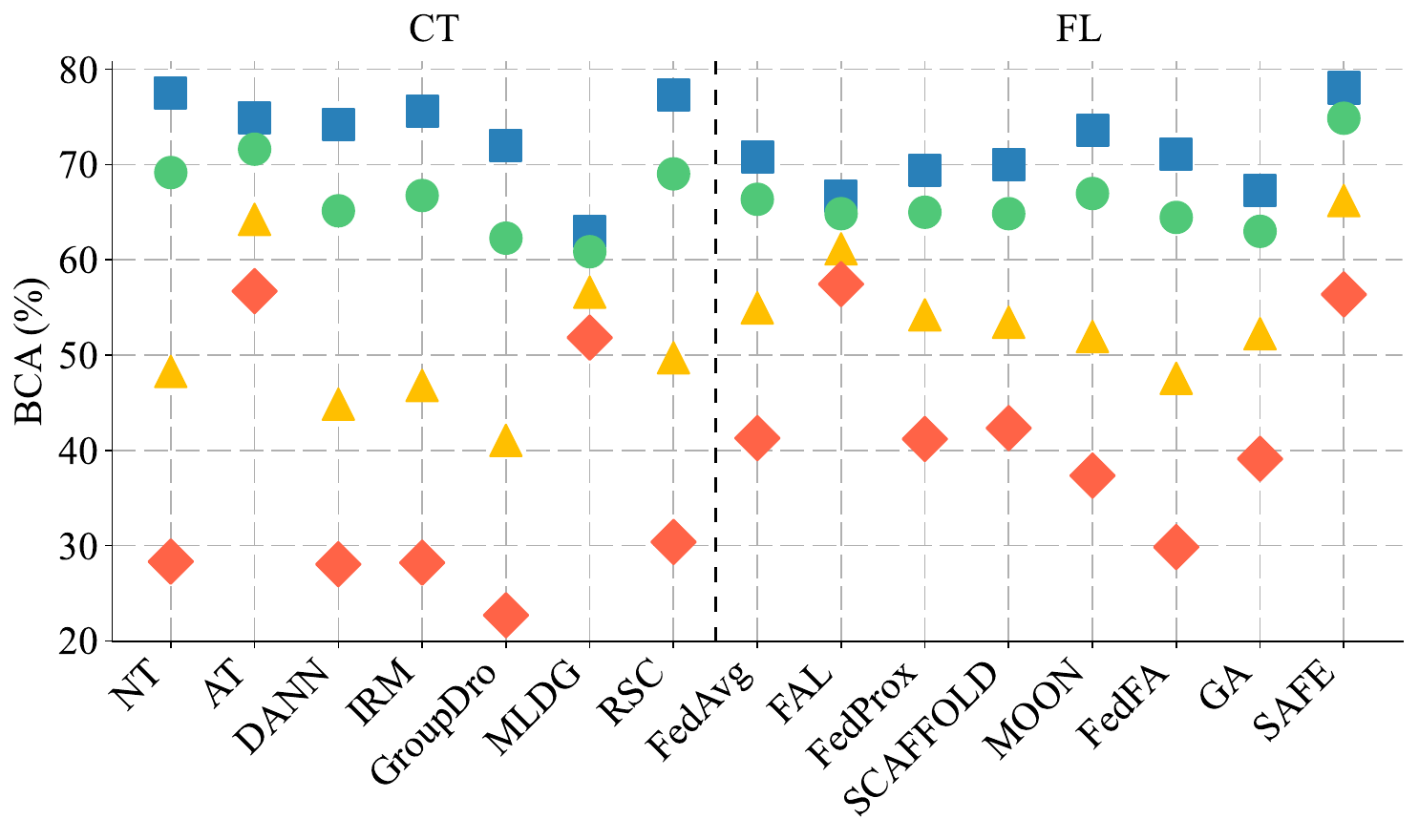}}
	\caption{Average BCAs on the adversarial samples under three white-box attacks with different magnitudes $\epsilon \in \{0, 0.01, 0.03, 0.05\}$ on the three MI datasets. (a) MI1; (b) MI2; and, (c) MI3.}	\label{fig:white_result_mi}
\end{figure}

\begin{figure}[htbp] 	\centering
	{\includegraphics[width=\linewidth]{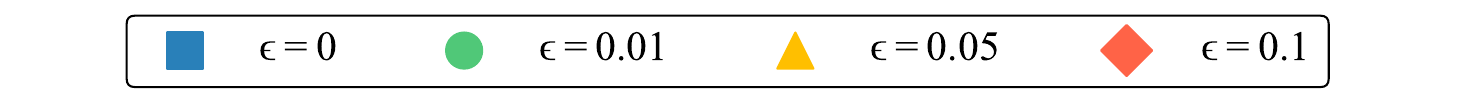}}
	\subfigure[MI1]{\label{fig:mi1_black}\includegraphics[width=\linewidth]{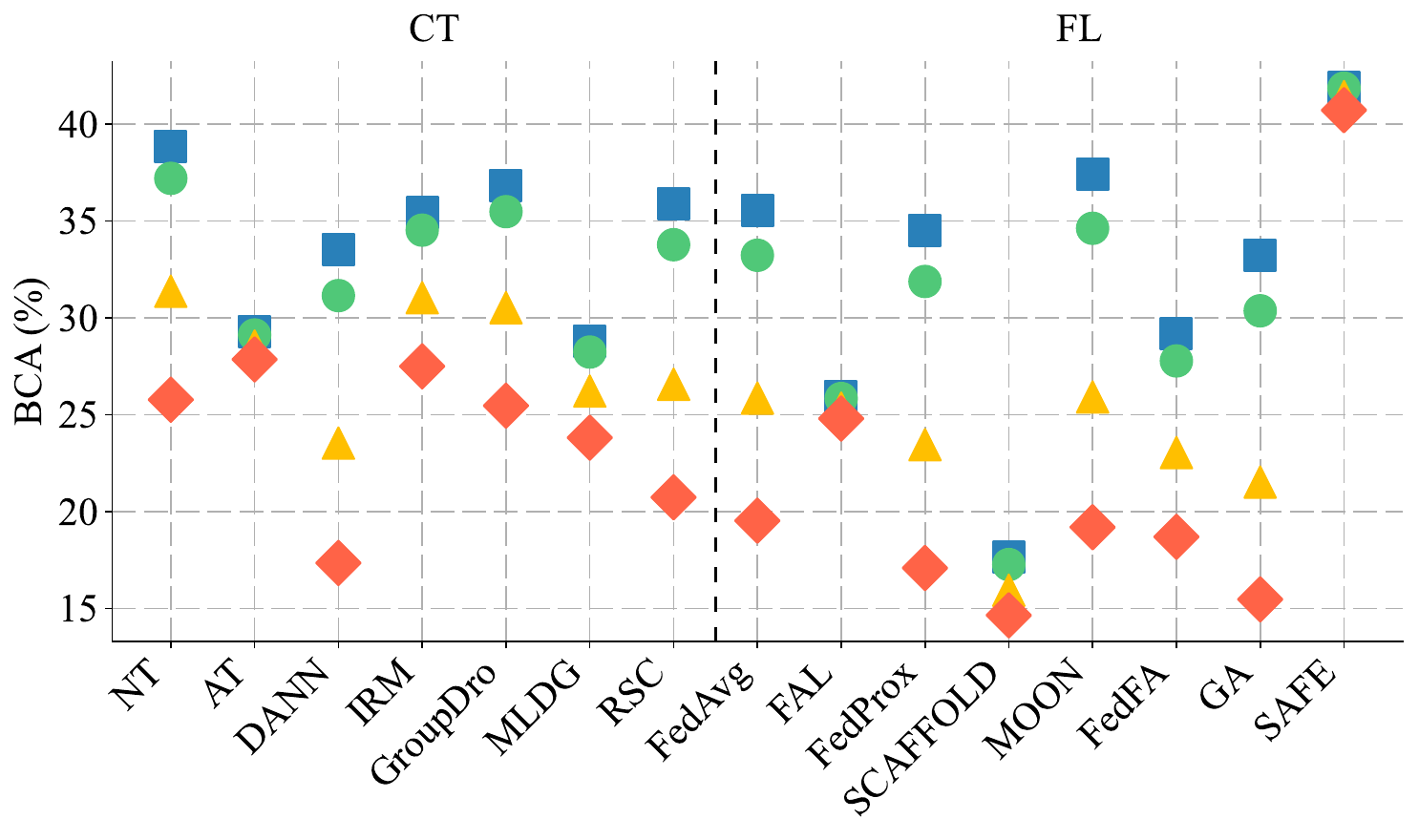}}
	\subfigure[MI2]{\label{fig:mi2_black}\includegraphics[width=\linewidth]{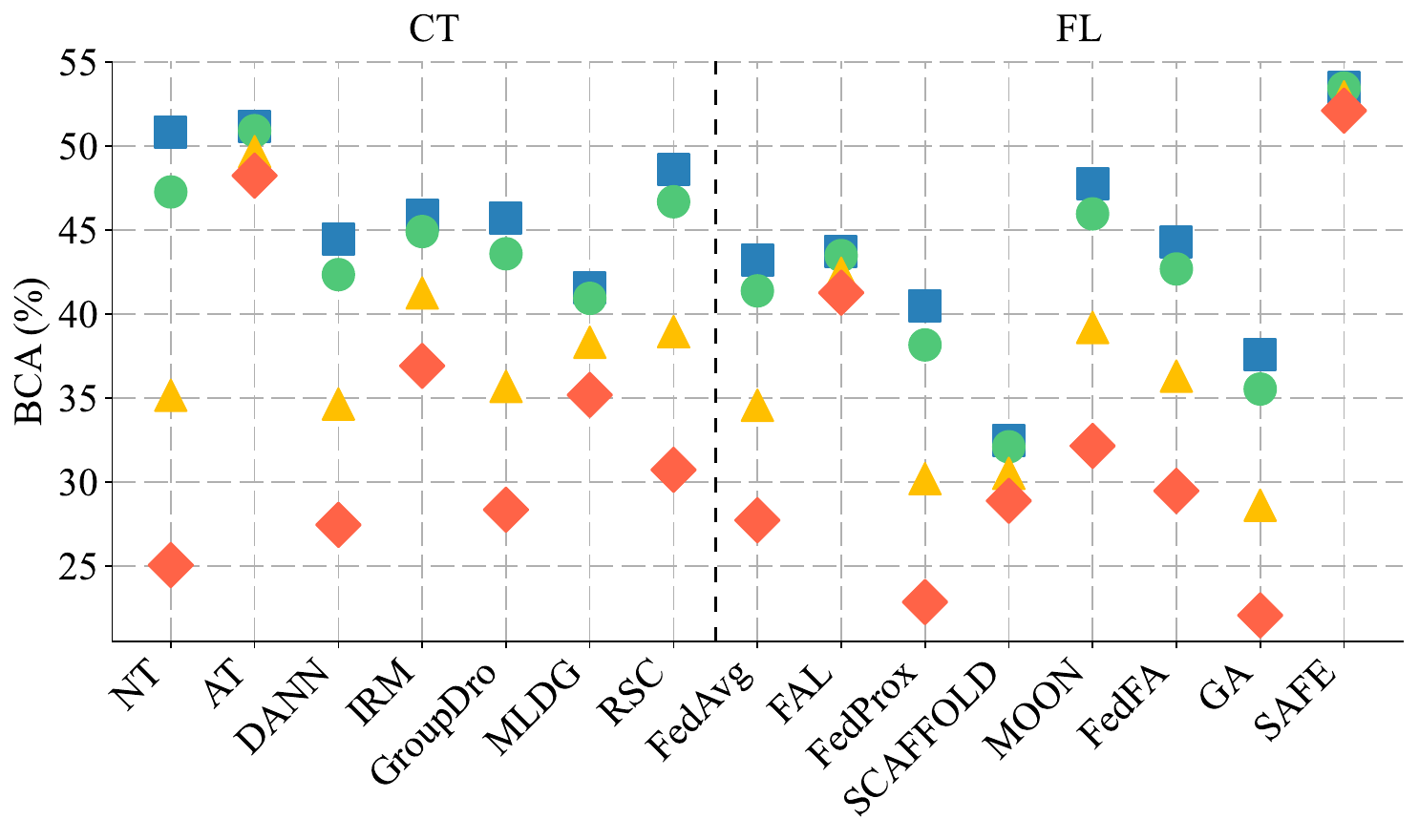}}
	\subfigure[MI3]{\label{fig:mi3_black}\includegraphics[width=\linewidth]{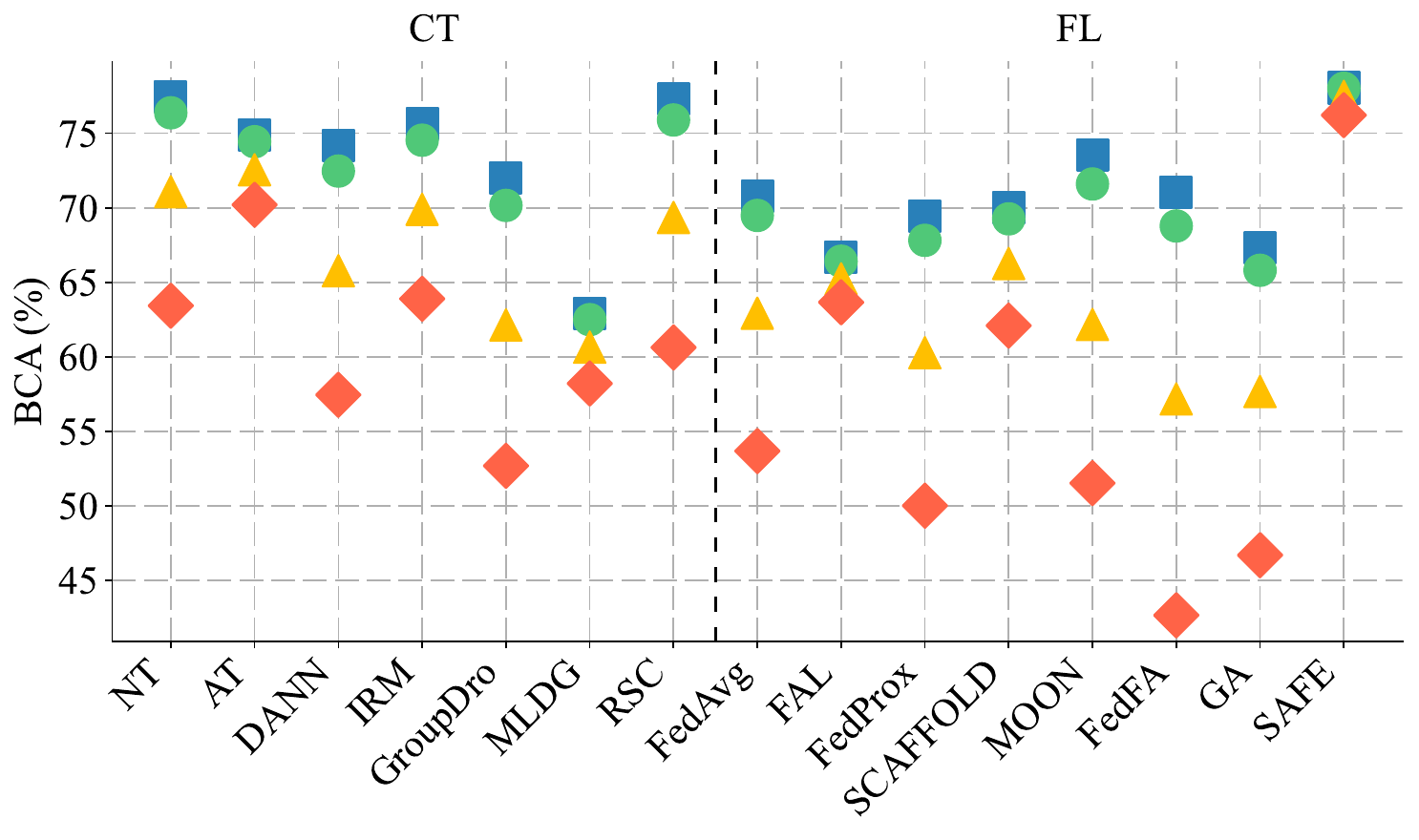}}
	\caption{Average BCAs on the adversarial samples under two black-box attacks with different magnitudes $\epsilon \in \{0,0.01,0.05,0.1\}$ on the three MI datasets. (a) MI1; (b) MI2; and, (c) MI3.}	\label{fig:black_result_mi}
\end{figure}

\begin{figure*}[htbp]	\centering
	{\includegraphics[width=0.5\linewidth]{legend_white-eps-converted-to.pdf}}\\
	\subfigure[ERP1]{\label{fig:erp1_white}\includegraphics[width=0.49\linewidth]{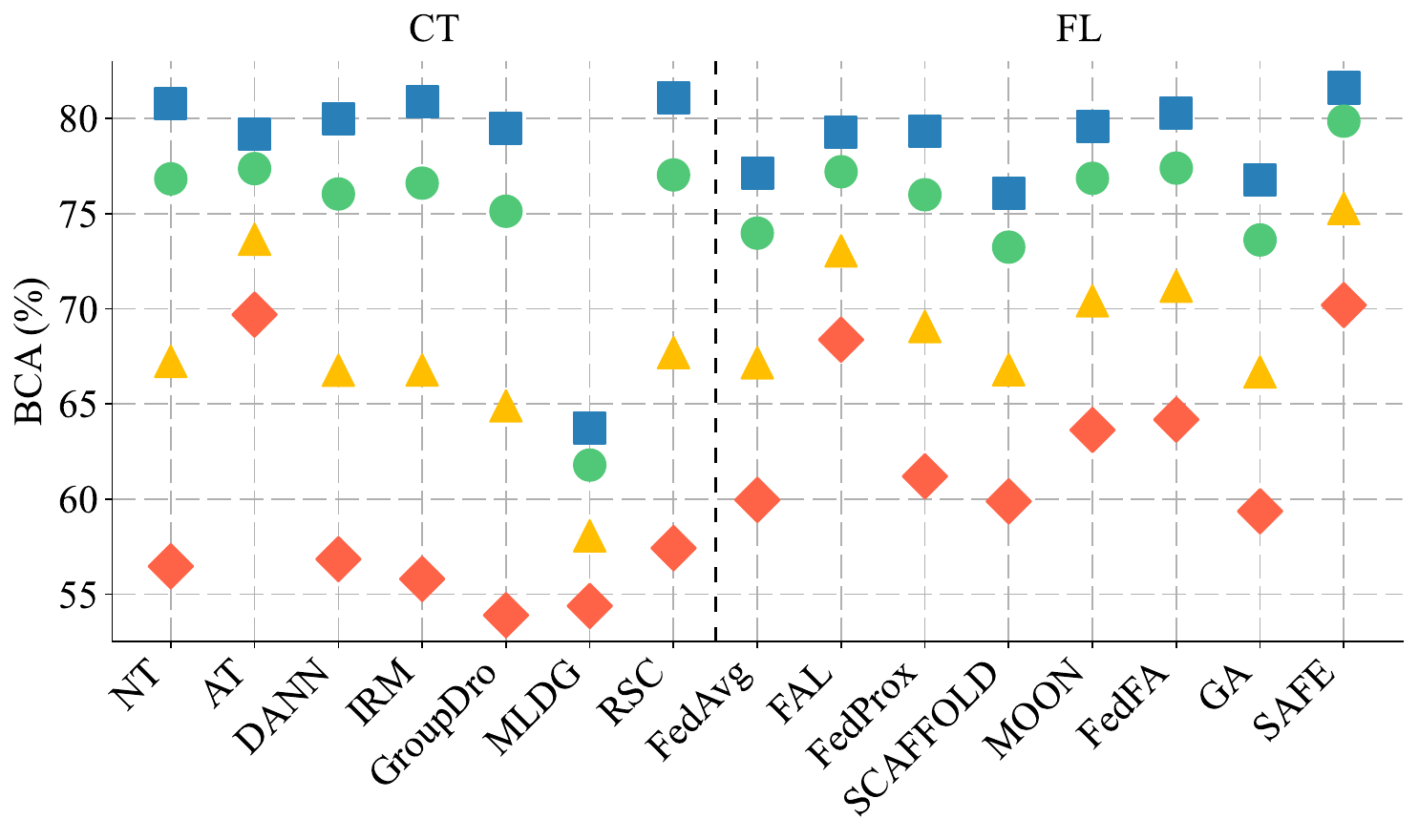}}
	\subfigure[ERP2]{\label{fig:erp2_white}\includegraphics[width=0.49\linewidth]{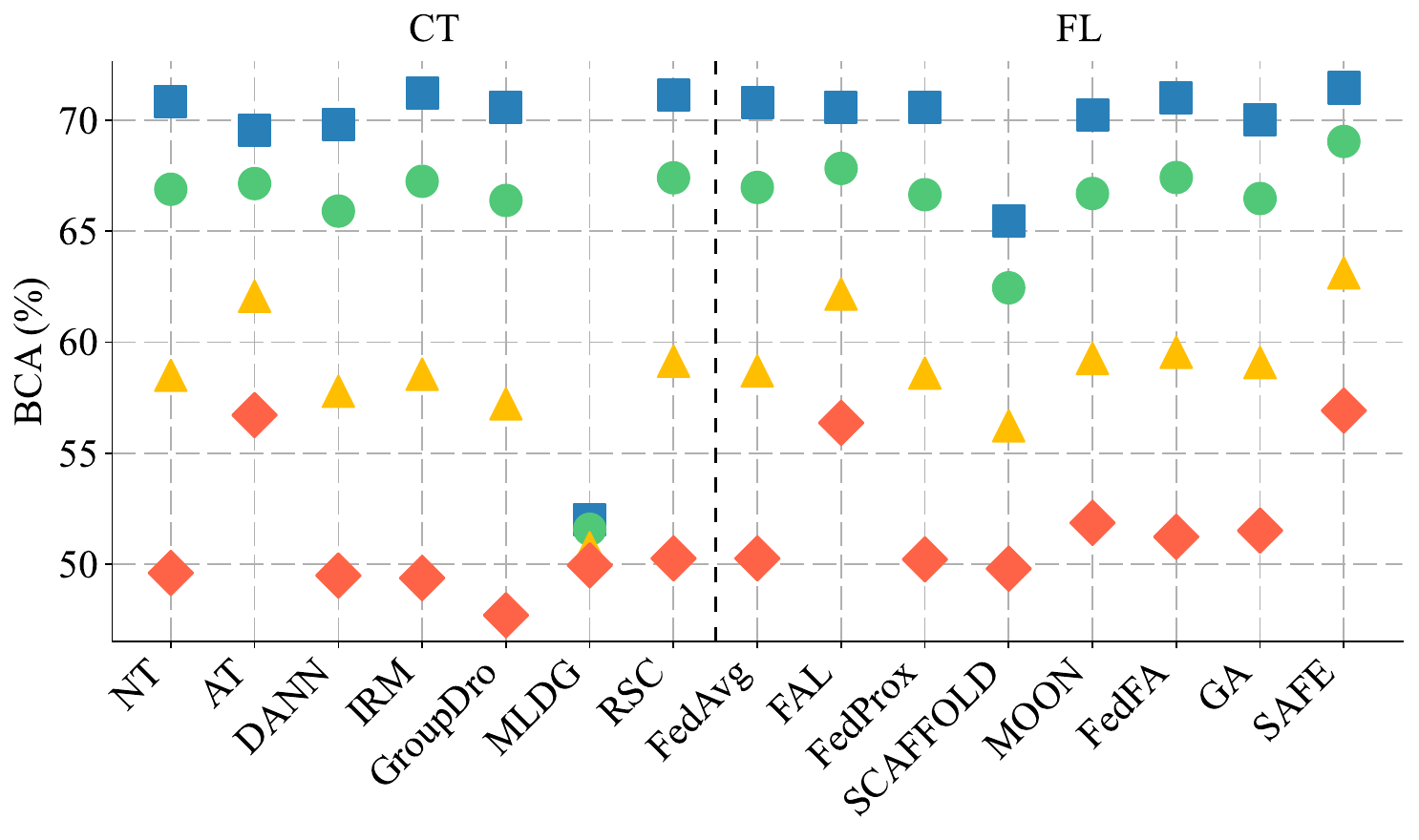}}
	\caption{Average BCAs on the adversarial samples under three white-box attacks with different magnitudes $\epsilon \in \{0, 0.01, 0.03, 0.05\}$ on the two ERP datasets. (a) ERP1; and, (b) ERP2.}	\label{fig:white_result_erp}
\end{figure*}

\begin{figure*}[htbp]	\centering
	{\includegraphics[width=0.5\linewidth]{legend_black-eps-converted-to.pdf}}\\
	\subfigure[ERP1]{\label{fig:erp1_black}\includegraphics[width=0.49\linewidth]{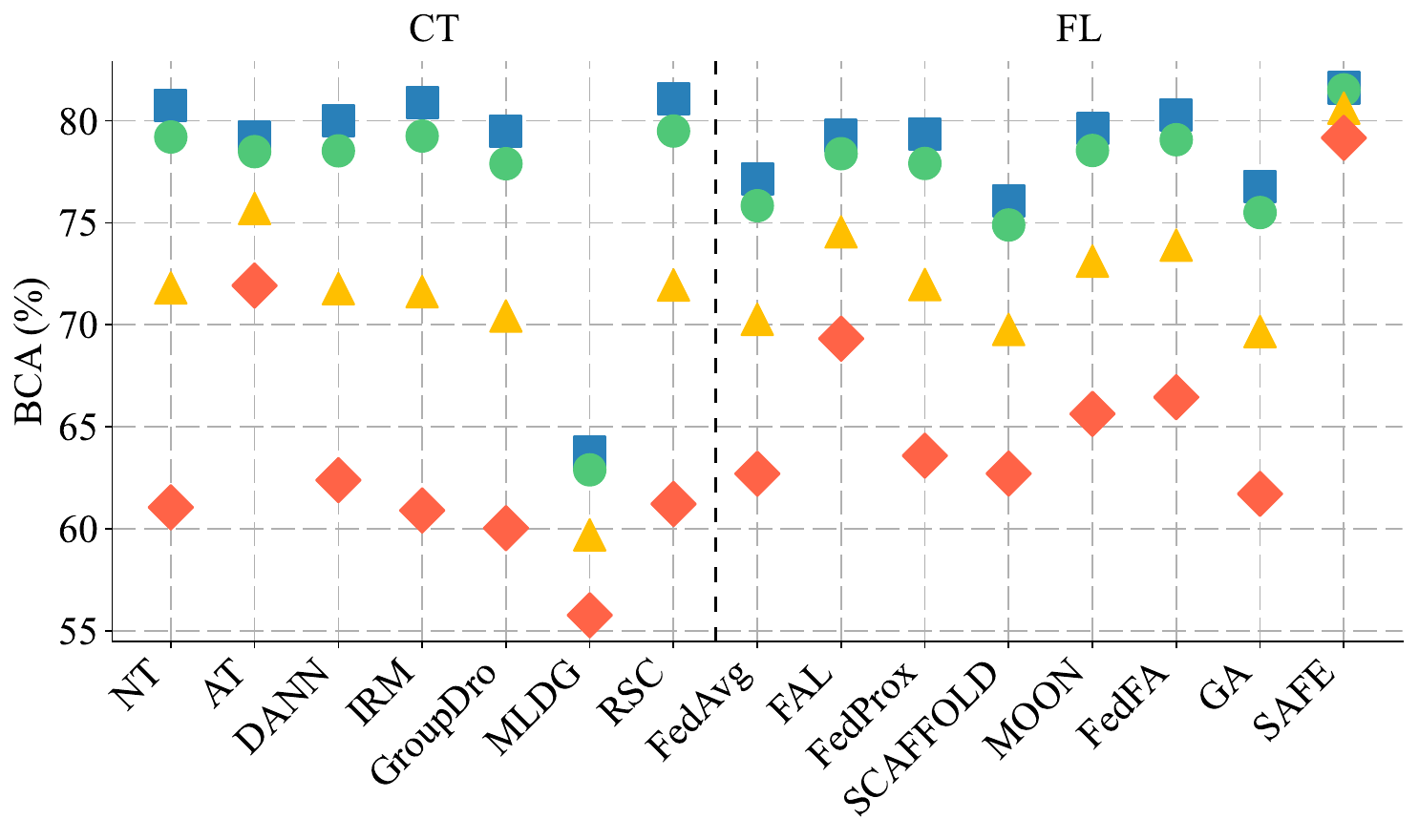}}	\subfigure[ERP2]{\label{fig:erp2_black}\includegraphics[width=0.49\linewidth]{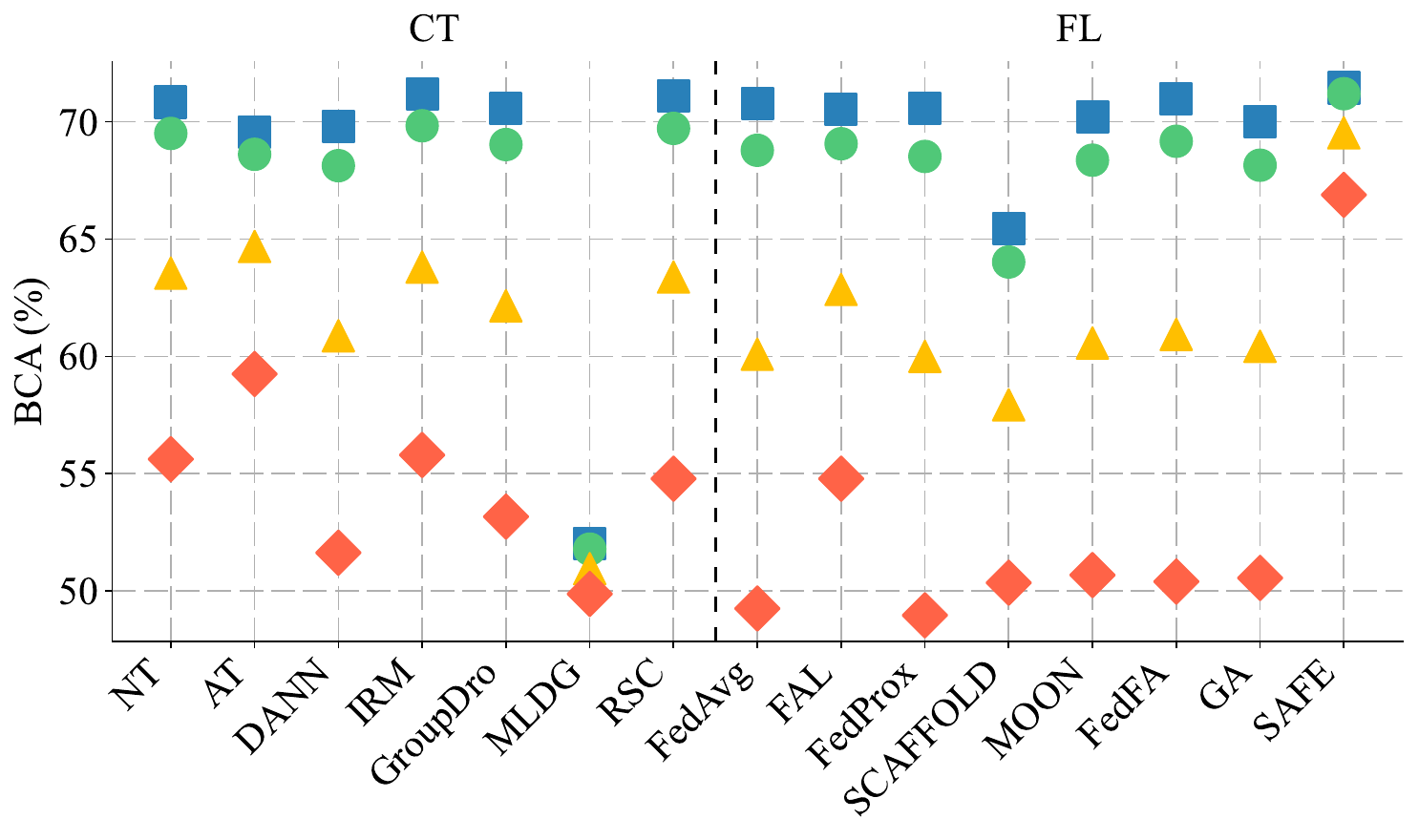}}
	\caption{Average BCAs on the adversarial samples under two black-box attacks with different magnitudes $\epsilon \in \{0,0.01,0.05,0.1\}$ on the two ERP datasets. (a) ERP1; and, (b) ERP2.}	\label{fig:black_result_erp}
\end{figure*}

Figs.~\ref{fig:white_result_mi}-\ref{fig:black_result_erp} respectively show the average BCAs on adversarial samples under three white-box attacks and two black-box attacks across the five datasets, with varying attack magnitude ($\epsilon=0$ corresponds to the benign samples).

Observe that:
\begin{enumerate}
\item In general, methods without adversarial defense mechanisms exhibited significant performance degradation on adversarial samples, particularly in white-box attacks. 	

\item Compared to CT, FL approaches had some ability to resist adversarial attacks (e.g., FedAvg outperformed NT under strong attacks), likely due to the inherent disturbance from client distribution heterogeneity in FL training.

\item Under white-box attacks, AT, FAL, and SAFE methods equipped with adversarial defense mechanisms demonstrated robust resilience against adversarial attacks, subject to only minor performance degradation.
    
\item Under black-box attacks, SAFE demonstrated negligible performance degradation as the attack intensity increased. This is significant in real-world applications, where attackers seldom have complete access to the model.
    
\item Overall, SAFE achieved the highest classification accuracy across diverse datasets, attack types, and intensity levels, demonstrating its effectiveness in defending against adversarial attacks while protecting EEG data privacy.
\end{enumerate}

\subsection{Ablation Studies}

Ablation studies were conducted to check if each strategy used in SAFE was effective and necessary. 

Table~\ref{tab:ablation} presents the average BCAs for benign samples, white-box adversarial samples, and black-box adversarial samples on MI1 and ERP1. While LBSN achieved competitive accuracy on benign samples, it showed severe vulnerability to white-box attacks. FAT or AWP alone significantly compromised the accuracy on benign samples. While using all three strategies may lead to a slight decrease in the accuracy on benign samples, it significantly enhanced the adversarial robustness. Overall, each strategy was effective, and using all three of them together resulted in the optimal performance.

\begin{table}[htbp]\centering \setlength{\tabcolsep}{2mm}
	\caption{Average BCAs (\%) for benign samples, white-box adversarial samples and black-box adversarial samples in ablation studies on MI1 and ERP1. The best accuracy in each column for each dataset is marked in bold.} \label{tab:ablation}
	\begin{tabular}{c|ccc|ccc|c}
		\toprule
		\multirow{2}{*}{Dataset} & \multicolumn{3}{c|}{Strategy} & \multicolumn{3}{c|}{Samples} & \multicolumn{1}{c}{\multirow{2}{*}{Avg.}} \\ \cmidrule{2-7}
		& LBSN   & FAT         & AWP         & Benign & White & Black & \multicolumn{1}{c}{} \\ \midrule
		\multirow{8}{*}{MI1}  & \XSolidBrush & \XSolidBrush & \XSolidBrush & 35.53  & 10.28 & 26.20 & 24.00                \\
		& \Checkmark   & \XSolidBrush & \XSolidBrush & \textbf{42.27}  & 12.94 & 40.85 & 32.02                \\
		& \XSolidBrush & \Checkmark   & \XSolidBrush & 26.11  & 20.27 & 25.51 & 23.96                \\
		& \XSolidBrush & \XSolidBrush & \Checkmark   & 35.59  & 13.95 & 26.60 & 25.38                \\
		& \Checkmark   & \Checkmark   & \XSolidBrush & 40.71  & 25.04 & 40.09 & 35.28                \\
		& \Checkmark   & \XSolidBrush & \Checkmark   & 41.89  & 15.64 & 40.59 & 32.71                \\
		& \XSolidBrush & \Checkmark   & \Checkmark   & 25.63  & 20.21 & 25.00 & 23.61                \\
		& \Checkmark   & \Checkmark   & \Checkmark   & 41.86  & \textbf{27.75} & \textbf{41.32} & \textbf{36.98}                \\ \midrule
		\multirow{8}{*}{ERP1} & \XSolidBrush & \XSolidBrush & \XSolidBrush & 77.15  & 67.04 & 69.60 & 71.26                \\
		& \Checkmark   & \XSolidBrush & \XSolidBrush & 81.23  & 70.95 & 79.47 & 77.22                \\
		& \XSolidBrush & \Checkmark   & \XSolidBrush & 79.19  & 72.62 & 73.88 & 75.23                \\
		& \XSolidBrush & \XSolidBrush & \Checkmark   & 80.16  & 70.43 & 71.96 & 74.18                \\
		& \Checkmark   & \Checkmark   & \XSolidBrush & 80.44  & 73.78 & 79.62 & 77.95                \\
		& \Checkmark   & \XSolidBrush & \Checkmark   & \textbf{81.66}  & 71.47 & 79.87 & 77.66                \\
		& \XSolidBrush & \Checkmark   & \Checkmark   & 79.93  & 73.53 & 74.72 & 76.06                \\
		& \Checkmark   & \Checkmark   & \Checkmark   & 81.62  & \textbf{75.11} & \textbf{80.44} & \textbf{79.06}                \\ \bottomrule
	\end{tabular}
\end{table}

\subsection{Impact of Client Selection and Local Optimization on FL Performance}

In practice, the configuration of FL requires careful consideration, particularly with respect to the number of selected clients per round ($m$) and the number of local optimization epochs ($E$). 

\begin{figure}[htbp]     \centering
	\includegraphics[width=\linewidth]{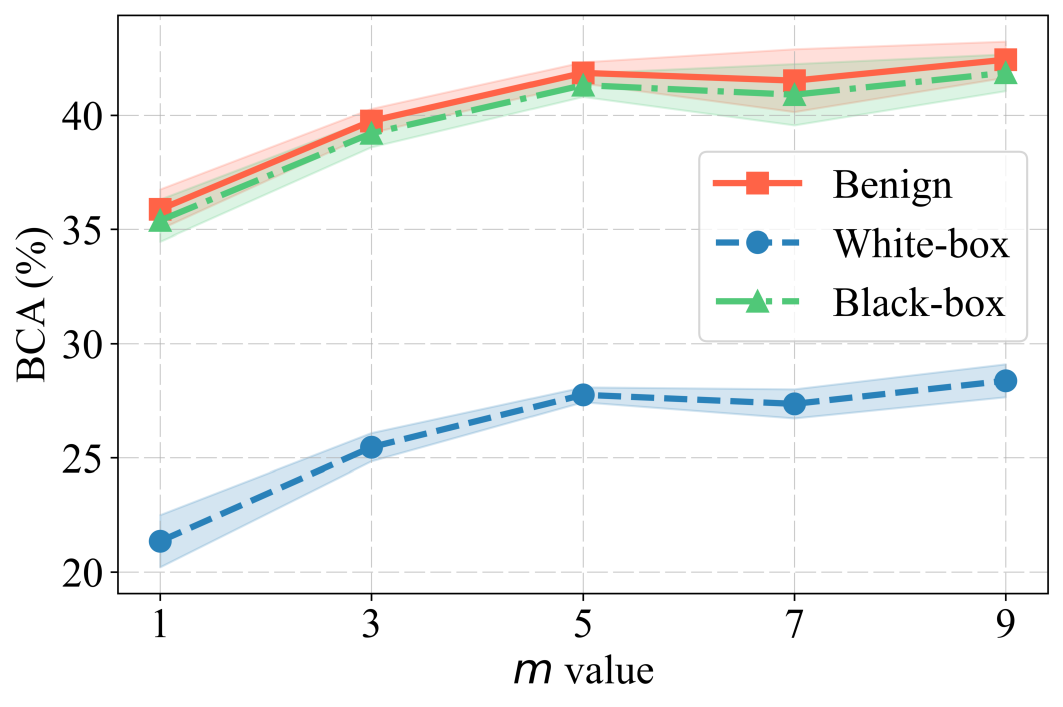}
	\caption{Average cross-subject BCAs for benign samples, white-box adversarial samples, and black-box adversarial samples under various values of $m$ (number of selected clients per round) on MI1.} 	\label{fig:mi1_m}
\end{figure}

\begin{figure}[htbp]     \centering
	\includegraphics[width=\linewidth]{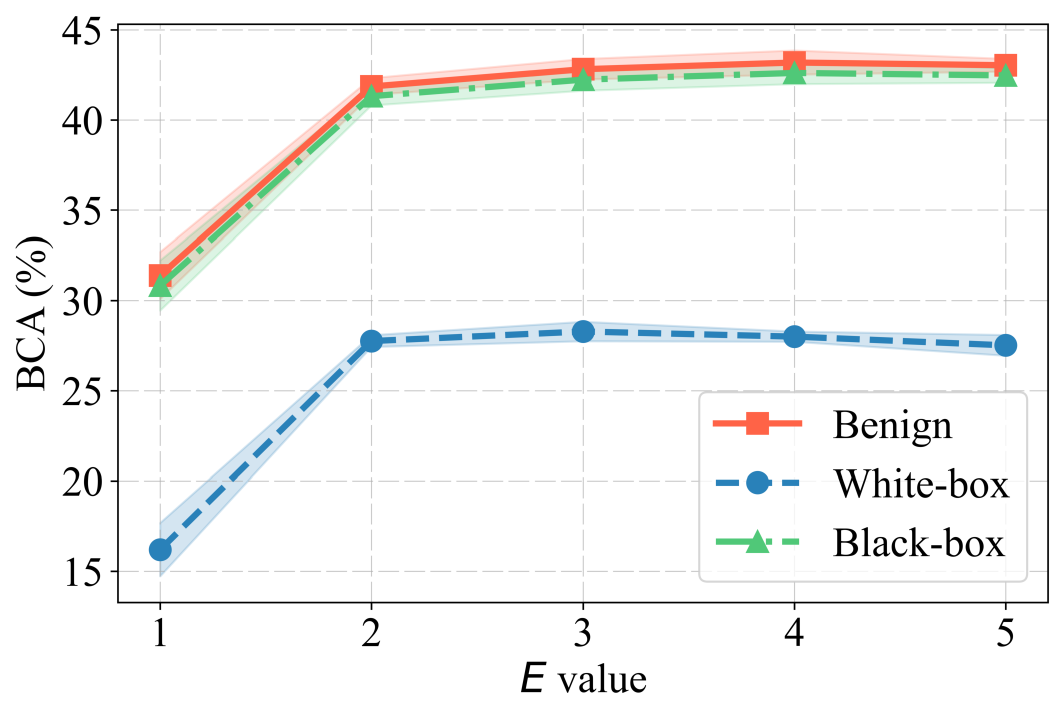}
	\caption{Average cross-subject BCAs for benign samples, white-box adversarial samples, and black-box adversarial samples under various values of $E$ (number of local optimization epochs) on MI1.}     \label{fig:mi1_e}
\end{figure}

Figs.~\ref{fig:mi1_m} and \ref{fig:mi1_e} respectively illustrate the average BCAs of SAFE under various values of $m$ and $E$ on MI1, for benign samples, white-box adversarial samples, and black-box adversarial samples. The performance of SAFE converged very fast for both $m$ and $E$, which is desirable in real-world BCIs.

\section{Conclusions} \label{sec:conclusions}

This paper has proposed SAFE to address three challenges in EEG-based BCIs: inadequate generalization, adversarial vulnerability, and privacy leakage. SAFE utilizes FL to protect user EEG data privacy, and LBSN to adapt to client data distribution shifts and improve generalization. Additionally, it adopts a dual-defense mechanism to defend against adversarial attacks. Experiments on five EEG datasets from MI and ERP paradigms demonstrated that SAFE outperformed 14 state-of-the-art approaches in terms of decoding accuracy and adversarial robustness, including CT methods that do not consider privacy protection at all. To our knowledge, SAFE is the first algorithm to simultaneously achieve high decoding accuracy, strong adversarial robustness, and reliable privacy protection, without using any calibration data from the target subject. These characteristics make it very desirable for real-world BCIs.

\bibliographystyle{IEEEtran}\bibliography{twjiabib}

\end{document}